\documentclass[12pt]{article}
\usepackage[margin=1in]{geometry}
\usepackage[utf8]{inputenc}
\usepackage[english]{babel}
\usepackage{amsmath, amssymb}
\usepackage{bbold}
\usepackage{amsthm} 
\usepackage{slashed}
\usepackage{physics}
\usepackage{mathtools} 
\usepackage{comment} 
\usepackage{setspace}
\usepackage[numbers,sort&compress]{natbib}
\usepackage{url}
\usepackage[all]{xy}
\usepackage{tikz-cd}
\usepackage{hyperref}
\usepackage{musicography}
\usepackage{lscape}
\hypersetup{
    colorlinks,
    citecolor=black,
    filecolor=black,
    linkcolor=black,
    urlcolor=black
}
\usepackage{mathrsfs}
\usepackage{float}

\DeclareMathOperator*{\argmin}{arg\,min}

\theoremstyle{definition}

\title{The Inverse of Exact Renormalization Group Flows as Statistical Inference}

\author{David S. Berman$^1$, Marc S. Klinger$^2$ \\
${}^1$ Centre for Theoretical Physics, Queen Mary University of London, Mile End Road, London E1 4NS \\
${}^{2}$ Department of Physics, University of Illinois, Urbana IL 61801, USA}

\begin{document}


\begin{flushright}
QMUL-PH-22-40
\end{flushright}
\bigskip
\bigskip

\centerline{\Large{The Inverse of Exact Renormalization Group Flows as Statistical Inference}}
\centerline{\Large{}}

\bigskip
\bigskip

\centerline{David S. Berman$^1$, Marc S. Klinger$^2$}
\bigskip
\bigskip

\centerline{${}^1$ Centre for Theoretical Physics, Queen Mary University of London, Mile End Road, London E1 4NS}
\centerline{${}^{2}$ Department of Physics, University of Illinois, Urbana IL 61801, USA}

\begin{abstract}
    We build on the view of the Exact Renormalization Group (ERG) as an instantiation of Optimal Transport described by a functional convection-diffusion equation. We provide a new information theoretic perspective for understanding the ERG through the intermediary of Bayesian Statistical Inference. This connection is facilitated by the Dynamical Bayesian Inference scheme, which encodes Bayesian inference in the form of a one parameter family of probability distributions solving an integro-differential equation derived from Bayes' law. In this note, we demonstrate how the Dynamical Bayesian Inference equation is, itself, equivalent to a diffusion equation which we dub \emph{Bayesian Diffusion}. Identifying the features that define Bayesian Diffusion, and mapping them onto the features that define the ERG, we obtain a dictionary outlining how renormalization can be understood as the inverse of statistical inference. 
\end{abstract}

\pagebreak

\tableofcontents

\pagebreak

\section{Introduction}

The Renormalization Group (RG) is used in physical settings to deduce how a theory changes when it is viewed at different \emph{{scales}}. In~Wilsonian RG, one regards the set of possible theories as a space coordinatized by a collection of coupling constants specifying all of the possible contributions to a classical action. An~RG flow can then be understood as a one-parameter family, or~\emph{{flow}}, in~the space of theories generated by a vector field on theory space referred to as the \emph{{Beta Function}} \cite{wilson1974renormalization}. The~flow is completely specified once the requisite initial data are provided, for~example, fixing a theory in the ultraviolet (UV) from which the RG flow begins ({
 by fixing a theory in the UV, we mean a theory that is a priori valid at all energy scales}).

A typical approach to RG is to deduce the beta function by sequentially coarse-graining the theory---that is, integrating out degrees of freedom that become suppressed at lower energy scales. In~field theoretic contexts, this approach makes use of perturbative techniques in order to perform the requisite functional integrals. From~a formal perspective, however, it is possible to study the properties of renormalization as an abstract flow equation without immediately concerning ourselves with any complicated or even intractable calculations that may be required to realize the flow explicitly. This approach to RG goes under the name of the \emph{Exact} Renormalization Group (ERG) and~will be the primarily focus of our~paper. 

In our approach to ERG, we shall regard a Quantum Field Theory (QFT) as equivalent to the specification of a probability distribution on the space of fields included in the theory, which we denote by $\mathfrak{F}(M)$ and refer to as the \emph{sample space} in standard probability theoretic nomenclature (to accommodate this perspective, we shall consider only Euclidean QFT). From~this perspective, the space of theories is isomorphic to the space of probability distributions on the sample space $\mathfrak{F}(M)$, denoted by $\mathcal{M}$. An~ERG flow is therefore equivalent to a one-parameter family of probability distributions on $\mathcal{M}$. 

From this point of view, we can still make contact with the Wilsonian picture by regarding a Wilsonian Effective Action, written in terms of a collection of coupling constants, as~specifying a \emph{{parameteric family}} of probability distributions. In~other words, Wilsonian RG corresponds to a particular coordinatization of the space $\mathcal{M}$. 

Following the lead of~\cite{polchinski1984renormalization,latorre2000exact,cotler2022renormalization} and others, it is natural to interpret the one-parameter family of probability distributions generated by an ERG flow as being governed by a functional \emph{{convection--diffusion}} equation on the sample space $\mathfrak{F}(M)$. Regarding ERG as a flow on the space of probability distributions over a given sample space contextualizes renormalization in a language that is amenable to applications outside of the usual realm of physical theory. In~particular, it suggests a manifestly information-theoretic interpretation for renormalization. Uncovering and presenting the details of this interpretation is the primarily objective of this~note. 

In seeking an answer to the question, ``What is the information-theoretic Interpretation of an ERG flow?'' we find it useful to consider a related question: ``What does it mean to \emph{{invert}} an ERG flow?''. Here, our understanding of ERG as being governed by a diffusion equation provides us with a direction. A~powerful method for formally inverting a diffusion process is \emph{{Bayesian Inversion}} \cite{dashti2017bayesian}. Bayesian Inversion is a probabilistic approach used to determine the initial data that was fed into a partial differential equation and subsequently generated an observed sequence of outcomes. As~the name implies, the~main tool employed in Bayesian Inversion is Bayes' Law. This lends some credence to the idea that Bayesian Inference may serve as an ``inverse process'' to~ERG. 

In our previous publication~\cite{berman2022dynamics}, we introduced \emph{{Dynamical Bayesian Inference}}, which recasts the Bayesian Inference as a dynamical system. Like ERG, Dynamical Bayesian Inference generates a one-parameter family of probability distributions satisfying a flow equation. Whereas in ERG, the~flow equation is deduced by sequential application of a coarse-graining law, in~Dynamical Bayes the flow equation is obtained by the sequential application of Bayes' law with respect to a continuously growing set of observed data. In~this respect, an~ERG flow continuously loses information ({hence, it is \emph{{diffusive}}}), while a Dynamical Bayesian flow continuously gains~information. 

In this note, we shall argue in favor of this interpretation. In~particular, we will demonstrate that the Dynamical Bayesian Flow equation gives rise to a convection--diffusion equation describing the evolution of an associated posterior predictive distribution \emph{{backwards}} against the collection of additional data. We refer to the process described by this equation as \emph{{Bayesian Diffusion}}, or~\emph{{Backward Inference}}. We interpret the equation governing Bayesian Diffusion as \emph{{defining}} an ERG flow, with~a coarse-graining procedure given by the continuous discarding of observed data. By~construction, this ERG is \emph{{inverted}} by the \emph{{forward}} Dynamical Bayesian inference flow that is obtained by reincorporating the lost data back into the model. Alternatively, starting from an ERG flow we identify the Dynamical Bayesian flow for which it is the Backward Inference process by drawing a correspondence between the partial differential equation governing ERG and the partial differential equation governing Bayesian diffusion. Ultimately, this correspondence suggests a fascinating information-theoretic interpretation for ERG: it can be regarded as the one-parameter family of probability distributions obtained by starting from a data-generating model and~continuously throwing away information in the form of observed~data. 

The organization of the paper is given as follows: in Section~\ref{ERG is Diffusion}, we review the formulation of ERG as a functional differential equation, specifically in the Wegner--Morris form. We also demonstrate that the Wegner--Morris equation is equivalent to a Fokker--Planck equation with a given potential function, establishing the correspondence between ERG and optimal transport. In~Section~\ref{SDEs and PDEs}, we introduce Stochastic Differential Equations (SDEs) and~discuss the relationship between SDEs and partial differential equations of the type appearing in ERG. In~Section~\ref{Dynamical Bayes}, we review Dynamical Bayesian Inference and~derive the Bayesian Diffusion equations. Finally, in~Section~\ref{ERG = Backwards Inference}, we establish the correspondence between Bayesian Diffusion and ERG explicitly. We conclude with Section~\ref{Discussion}, in which we review our findings and~discuss future research~directions.

\section{ERG Equation as a Functional Diffusion~Equation} \label{ERG is Diffusion}

In this section, we will provide an overview of functional renormalization with the objective of demonstrating how ERG can be understood as a functional differential equation. Our presentation will only focus on the aspects of ERG that are relevant for this work; for~a more complete overview, see~\cite{polchinski1984renormalization,cotler2022renormalization,bagnuls2001exact,rosten2012fundamentals} and the references therein. The~presentation here follows very closely with that in~\cite{cotler2022renormalization}. 

Let us consider a theory with fields $\Phi \in \mathfrak{F}(M)$. Here, we are using a notation in which capital letters correspond to random variables, and~lower case letters to realizations. In~this case, $\Phi$ is to be regarded as a random variable taking values in a space of continuous functions on a manifold $M$. The~distribution over $\Phi$ is a probability functional \mbox{$P[\phi(x)] \propto e^{-S[\phi(x)]}$ where $S[\phi(x)]$} is the Euclidean action. The~idea of RG is that we are only capable of probing scales less than a cutoff $\Lambda$. As~we change our cutoff, we obtain a family of probability distributions, $P_{\Lambda}[\phi(x)] \propto e^{-S_{\Lambda}[\phi(x)]}$, corresponding to effective descriptions for the field $\Phi$ at each measurement threshold. ERG provides a description for the changes to the effective probability distribution in the form of a flow equation:
\begin{equation} \label{General ERG}
    -\Lambda \frac{d}{d\Lambda} P_{\Lambda}[\phi] = \mathcal{F}\left[P_{\Lambda}[\phi], \frac{\delta P_{\Lambda}[\phi]}{\delta \phi}, \frac{\delta^2 P_{\Lambda}[\phi]}{\delta \phi \delta \phi}, ...\right] \, .
\end{equation}
where $\mathcal{F}$ is a functional of $P_{\Lambda}[\phi]$ and all its functional derivatives and specifies a particular ERG \emph{scheme}. 

\subsection{Polchinski's~Equation}

The canonical example of ERG is Polchinski's approach~\cite{polchinski1984renormalization}. We begin with the partition function of a scalar field with source $J$ given by:

\begin{equation}
    Z_{\Lambda}[J] := \int \mathcal{D}\phi \; \exp{-\frac{1}{2} \int \frac{d^d p}{(2\pi)^d} \left(\phi(p) \phi(-p) (p^2 + m^2) K^{-1}_{\Lambda}(p^2) + J(p) \phi(-p)\right) - S_{int,\Lambda}[\phi]} \, .
\end{equation}
Here, $S_{int,\Lambda}$ is the interacting action, and~$K_{\Lambda}(p^2)$ is a cutoff function that differentially weighs modes $\phi(p)$ based on their~momenta. 

Polchinski's idea was to consider a scale $\Lambda_R < \Lambda$ and integrate out modes down to $\Lambda_R$. To~this end, we assume that $J(p)$ has compact support in the sphere of momenta with radius $\Lambda_R - \epsilon$ for a small $\epsilon > 0$. If~$\Lambda_R$ is only infinitesimally smaller than $\Lambda$, we can compute the differential change to $Z_{\Lambda}[J]$ on account of integrating out the shell of modes between $\Lambda$ and $\Lambda_R$. Then, we demand that:
\begin{equation} \label{RG condition}
    -\Lambda \frac{d}{d\Lambda} Z_{\Lambda}[J] = A_{\Lambda} Z_{\Lambda}[J] \, .
\end{equation}
where $A_{\Lambda}$ is a constant for each value of $\Lambda$. If~(\ref{RG condition}) holds, any correlation functions below the changed scale (i.e., for which one can take functional derivatives with respect to $J$) will be \emph{unchanged}. Hence, this form of RG respects the fact that measured correlation functions ought to be independent of the RG flow below the relevant momentum~scale. 

Expanding out (\ref{RG condition}), we find:

\begin{equation}
    -\Lambda \frac{d}{d\Lambda} Z_{\Lambda}[J] = \int \mathcal{D} \phi \left( \frac{1}{2} \int \frac{d^dp}{(2\pi)^d} \phi(p)\phi(-p)(p^2 + m^2) \Lambda \frac{\partial K^{-1}_{\Lambda}(p^2)}{\partial \Lambda} + \Lambda \frac{\partial S_{int,\Lambda}[\phi]}{\partial \Lambda}\right) e^{-S_{\Lambda}[\phi,J]} \, .
\end{equation}
Again, $K_{\Lambda}(p^2)$ is a function with prescribed $\Lambda$ dependence. The~behavior governed by (\ref{RG condition}) is that of $S_{int,\Lambda}$. Polchinski showed that one can consistently satisfy (\ref{RG condition}) by taking $S_{int,\Lambda}$ to satisfy the functional differential equation:

\begin{equation} \label{Polchinski Equation}
    -\Lambda \frac{\partial S_{int,\Lambda}[\phi]}{\partial \Lambda} = \frac{1}{2} \int d^dp (2\pi)^d (p^2 + m^2)^{-1} \Lambda \frac{\partial K_{\Lambda}(p^2)}{\partial \Lambda} \left\{\frac{\delta^2 S_{int,\Lambda}}{\delta \phi(p) \delta \phi(-p)} - \frac{\delta S_{int,\Lambda}}{\delta \phi(p)} \frac{\delta S_{int,\Lambda}}{\delta \phi(-p)}\right\} \, .
\end{equation}

Following the approach of (\ref{General ERG}), we define the probability functional: $P_{\Lambda}[\phi] = e^{-S_{\Lambda}[\phi]}/Z_{\Lambda}$, which is explicitly the probability distribution for the field $\Phi$ at scale $\Lambda$. Then, we can truly write the Polchinski equation in the form of a convection--diffusion equation as:

\vspace{-12pt}
\begin{multline} \label{Polchinski Heat Equation}
    -\Lambda \frac{d}{d\Lambda} P_{\Lambda}[\phi] = \frac{1}{2} \int d^d p (2\pi)^d (p^2 + m^2)^{-1} \Lambda \frac{\partial K_{\Lambda}(p^2)}{\partial \Lambda} \frac{\delta^2}{\delta \phi(p) \delta \phi(-p)} P_{\Lambda}[\phi] \\
    + \frac{1}{2} \int d^d p (2\pi)^d (p^2 + m^2)^{-1} \Lambda \frac{\partial K_{\Lambda}(p^2)}{\partial \Lambda} \frac{\delta}{\delta \phi(p)} \left( \frac{2(p^2 + m^2)}{(2\pi)^d K_{\Lambda}(p^2)} \phi(p) P_{\Lambda}[\phi]\right) \, .
\end{multline}
Throughout this note, we shall emphasize the sense in which such equations can be understood as functional analogs of more well defined finite dimensional equations. For~example, (\ref{Polchinski Heat Equation}) ought to be compared to the finite dimensional equation:
\begin{equation}
    \frac{d}{dt} p_t(y) = \sum_{i,j} g^{ij} \frac{\partial}{\partial y^i} \frac{\partial}{\partial y^j} p_t(y) + \sum_{i,j}  \frac{\partial}{\partial y^i}(g^{ij} v_j(y) p_t(y)) \, ,
\end{equation}
where we have written sums explicitly to make the analogy clear. To~move between the finite dimensional equation and the Polchinski equation, we have made use of the dictionary found in Table \ref{Finite Dimensional Diffusion and Polchinski's Equation}
.

\subsection{Wegner--Morris~Flows}

We have now established that Polchinski's equation is simply an infinite dimensional convection--diffusion equation. It can be shown that a family of such equations exists that satisfies the constraint (\ref{RG condition}). This family is defined by considering different \emph{{choices}} for the metric and drift velocity appearing in {Table}~\ref{Finite Dimensional Diffusion and Polchinski's Equation}. The~choice of this data then corresponds to a \emph{{choice of scheme}} for the ERG flow. The~aforementioned family of ERG schemes is given explicit representation in terms of the Wegner--Morris flow equation~\cite{wegner1973renormalization,wegner1974some,morris1994exact,morris1994derivative,morris1998elements}. As~a one-parameter family of probability distributions, the~Wegner--Morris flow is governed by the~equation:
\begin{equation} \label{WM Flow}
    -\Lambda \frac{d}{d\Lambda} P_{\Lambda}[\phi] = \int d^d x \frac{\delta}{\delta \phi(x)}\left(\Psi_{\Lambda}[\phi,x] P_{\Lambda}[\phi]\right) \, .
\end{equation}

\begin{table}[H]
\centering
\begin{tabular}{llll}
                                           & $\textbf{Finite Dim.}$ & $\textbf{Infinite Dim.}$                                                                                                                                                                                                                                        \\
$\textbf{Time Parameter}$                  & $t$                          & $\ell = \ln(\Lambda)$                                                                                                                                                                                                                                                         \\
$\textbf{Random Variable}$                          & $y^i$ & $\phi(p)$ \\

$\textbf{Metric}$                          & $g^{ij}$                     & $\dot{C}_{\Lambda}(p) = (2\pi)^d(p^2 + m^2)^{-1} \Lambda \frac{\partial K_{\Lambda}(p^2)}{\partial \Lambda}$                                                                                                                                                                                                                                        \\
$\textbf{Drift Velocity}$                       & $g^{ij} v_j$                          & $\Psi[\phi] = \int d^dp \; \dot{C}_{\Lambda}(p) \frac{2(p^2 + m^2)}{(2\pi)^d K_{\Lambda}(p^2)} \phi(p)$                                                                                                                                                                                                                                               
                                                                                                                                                                                            
\end{tabular}
\caption{Dictionary relating Finite Dimensional Diffusion and Polchinski's ERG equation.}
\label{Finite Dimensional Diffusion and Polchinski's Equation}
\end{table}

The Wegner--Morris \emph{scheme} is encapsulated in the kernel $\Psi_{\Lambda}[\phi,x]$. Again, it is useful to compare to a finite dimensional flow equation:
\begin{equation} \label{Wegner--Morris}
    \frac{d}{dt} p_t(y) = \sum_i \frac{\partial}{\partial y^i}\left(V^i(p_t,y) p_t\right) \, .
\end{equation}
Here, $\Psi_{\Lambda}$ has been replaced by a vector field that depends not only on $y$ but on the entire probability function $p_t$. It is often natural to regard $V$ as the gradient of a scalar function $W(p_t, y)$, which also depends on $p_t$, i.e.,~$V^i = g^{ij}(y) \frac{\partial}{\partial y^j} W(p_t,y)$. Then, we can write:
\begin{equation}
    \frac{d}{dt} p_t(y) = \sum_{ij} \frac{\partial}{\partial y^i}\left(g^{ij}(y) \frac{\partial}{\partial y^j} W(p_t,y) p_t\right) \, .
\end{equation}

A natural interpretation of (\ref{WM Flow}) is that it simply reparameterizes the field at each new scale. To~be precise, as $\Lambda$ changes we find:
\begin{equation} \label{Field Reparameterization}
    \phi'(x) = \phi(x) + \frac{\delta \Lambda}{\Lambda} \Psi_{\Lambda}[\phi,x]  \, .
\end{equation}
This implies that the Wegner--Morris flow conserves probability; it only changes the way that the probability is distributed. This is the reason why the Wegner--Morris family of flow equations satisfies (\ref{RG condition}). On~account of this new interpretation, $\Psi_{\Lambda}$ is given the title of the \emph{{Reparameterization Kernel}}. 

Following the intuition from the finite dimensional case, we can represent the Reparameterization Kernel as a functional gradient:
\begin{equation}
    \Psi_{\Lambda}[\phi,x] = - \int d^d y \frac{1}{2} \dot{C}_{\Lambda}(x,y) \frac{\delta \Sigma_{\Lambda}[\phi]}{\delta \phi(y)} \, .
\end{equation}
Here, $\dot{C}_{\Lambda}(x,y)$ is playing the same role it did in the Polchinski equation as an inverse metric on the sample space of fields; however, we have not yet fixed its functional form, and, indeed, it may differ from Polchinski's choice. In~the literature, $\dot{C}_{\Lambda}(x,y)$ is called the ERG~kernel. 

A popular choice for $\Sigma_{\Lambda}$ is given by the difference between the renormalized action of the theory and a second action, $\hat{S}_{\Lambda}$, called the \emph{seed action}:
\begin{equation} \label{Seed Action}
    \Sigma_{\Lambda}[\phi] := S_{\Lambda}[\phi] - 2\hat{S}_{\Lambda}[\phi]  \, .
\end{equation}
The factor of two is conventional. In~this framework, an~RG scheme is therefore specified entirely by a choice of $\dot{C}_{\Lambda}$ and $\hat{S}_{\Lambda}$. For~example, we can reconcile the Polchinski equation from the Wegner--Morris set up by taking:

\begin{equation}
    \dot{C}_{\Lambda}(p^2) = (2\pi)^d (p^2 + m^2)^{-1} \Lambda \frac{\partial K_{\Lambda}(p^2)}{\partial \Lambda}; \;\;\;\;\; \hat{S}_{\Lambda}[\phi] = \frac{1}{2}\int \frac{d^d p }{(2\pi)^d} (p^2 + m^2) K^{-1}_{\Lambda}(p^2) \phi(p) \phi(-p)   \, .
\end{equation}

\subsection{Field Reparameterization and Scheme~Independence} \label{Morris Scheme}

In Quantum Field Theory, fields are not an observable quantity but~rather a device used to encode a theory. In~this respect, we should regard field parametrization as a choice of coordinates on the sample space of the theory and~require that the theory be invariant under diffeomorphisms of these coordinates. Following this line of thought, the author of~\cite{latorre2000exact} showed that the field reparamterization appearing in (\ref{Field Reparameterization}) can be interpreted as a gauge redundancy, and, as~a consequence, the~drift component of the Wegner--Morris equation can be interpreted as a \emph{{choice of gauge}}. This perspective can be made concrete by viewing the convective derivative of the resulting flow equation as a covariant derivative, with~the vector field generating the drift playing the role of a gauge field. Changing the drift field changes the description of the ERG only cosmetically; in~particular, the expectation values of relevant operators are left-invariant under a change of gauge. We shall adopt this perspective and regard the prescription of drift in an ERG as tantamount to a \emph{{choice of scheme}}.

\subsection{Wegner--Morris and~Fokker--Planck} \label{WM = FP}

In the previous subsection, we saw that ERG can be understood as a functional differential equation associated with the Wegner--Morris Equation~(\ref{Wegner--Morris}). In~this section, we shall demonstrate how the Wegner--Morris equation may be regarded as a form of the Fokker--Planck equation. For~simplicity, we shall work in the context of probability theory on a differentiable sample space, $S$. 

Given a function $F: M \rightarrow \mathbb{R}$ on a (psuedo)-Riemannian manifold $M$ with metric $g$, we define the gradient as follows: for any path $\gamma: [0,1] \rightarrow M$, with~$\gamma(0) = x$, the~gradient of $F$ at $x$ with respect to the metric $g$ is the tangent vector $\text{grad}_g F(x)$ such that:
\begin{equation}
    \frac{d}{dt} \gamma^* F \bigg\rvert_{t = 0} = g(\text{grad}_g F(x), \gamma_* \frac{d}{dt})\bigg\rvert_{t = 0} \, .
\end{equation}

An equivalent definition that does not require the introduction of a path simply defines the gradient in terms of the exterior derivative on $M$:
\begin{equation} \label{Definition of Gradient}
    dF(X) = g(\text{grad}_g F, X); \;\;\; \text{grad}_g F = (dF)_{\musSharp} \, .
\end{equation}
Here, the~map $\musSharp: T^*M \rightarrow TM$ corresponds to the usual notion of index raising via the metric, mapping one form into vectors
. To~be precise, given $\alpha, \beta \in T^*M$ we have: $\alpha_{\musSharp}(\beta) = g^{-1}(\alpha, \beta)$. 

Let $\mathcal{M} = \text{dens}(S)$ denote the manifold of probability distribution functions on the sample space $S$. The~tangent space to $\mathcal{M}$ at a point $p$ is defined by:
\begin{equation} \label{Tangent to Space of Densities}
    T_p \mathcal{M} := \left\{\eta \in C^{\infty}(S) \; | \; \int_S \text{Vol}_S \; \eta = 0\right\} \, .
\end{equation}
This is a good definition since functions $\eta \in T\mathcal{M}$ can be regarded as perturbations to probability densities that do not spoil the integral normalization ({$\star$ is the Hodge star on $S$}):
\begin{equation}
    \int_S \star(p + \eta) = \int_S \star p + \int_S \star \eta = \int_S \star p = 1   \, .
\end{equation}
Hence, perturbations $p \mapsto p + \eta$ still belong to the manifold $\mathcal{M}$. 

Now, as~a first exercise in understanding gradients on $\mathcal{M}$, let us consider the most straightforward combination of metric and scalar function on this manifold. $\mathcal{M}$ is naturally equipped with an $\ell^2$ metric of the form
\begin{equation}
    G_{\ell^2}(\eta_1, \eta_2) := \int_S \eta_1 \wedge \star \eta_2  \, .
\end{equation}
Moreover, there is a natural functional to consider, namely, the \emph{{Dirichlet Energy Functional}} $\mathcal{E}: \mathcal{M} \rightarrow \mathbb{R}$ defined by ({$d$ is the exterior derivative on $S$ \emph{not} on $\mathcal{M}$}):
\begin{equation}
    \mathcal{E}[p] = \frac{1}{2} \int_S dp \wedge \star dp   \,  .
\end{equation}

The exterior derivative on $\mathcal{M}$ should be understood as the variational derivative with respect to the density $p$. Thus, by~standard techniques, we can compute:
\begin{equation}
    \delta \mathcal{E}[p] = \int_S d\delta p \wedge \star dp = \int_S (-1)^{d} \delta p \wedge d \star dp + \int_S d(\delta p \wedge \star dp)   \, .
\end{equation}
A probability distribution must necessarily have bounded support to satisfy an integral normalization condition; hence, in this and any future computations we can safely take the boundary variation to zero. Thus, we arrive at the desired result:
\begin{equation}
    \delta \mathcal{E}[p] = \int_S (-1)^d \delta p \wedge d \star dp = G_{\ell^2}((-1)^d \star d \star d p, \delta p)   \, .
\end{equation}
Matching this to the definition of the gradient (\ref{Definition of Gradient}), we conclude that the gradient of the Dirichlet Energy Functional with respect to the $\ell^2$ metric on $\mathcal{M}$ is equivalent to the Laplacian of $p$:
\begin{equation}
    \text{grad}_{\ell^2} \mathcal{E}[p] = - \Delta p   \, .
\end{equation}
An immediate corollary of this fact is that it allows us to write the standard heat equation, $\frac{\partial}{\partial t} p_t = \Delta p_t$, in~the form of a gradient flow as
\begin{equation}
    \frac{\partial}{\partial t} p_t = -\text{grad}_{\ell^2} \mathcal{E}[p_t]  \, .
\end{equation}

Remarkably, if~we choose a different metric for the space of probability distributions $\mathcal{M}$, we can reconcile the heat equation as a gradient flow with respect to a \emph{{different}} functional of the probability distribution. Of~particular interest to us is the task of constructing a metric for which the potential is the differential entropy:
\begin{equation}
    S[p] = - \int_S p \wedge \star \ln(p)        \, .
\end{equation}

To find such a metric, we begin by establishing the following isomorphism of $T\mathcal{M}$: given $\eta \in T_p \mathcal{M}$ and $\hat{\eta} \in C^{\infty}(S)$, the following equation implicitly defines an isomorphism between them, up~to a constant:
\begin{equation} \label{Tangent Space Isomorphism}
    \eta = \star d i_{p (d\hat{\eta})_{\musSharp}} \text{Vol}_S   \, .
\end{equation}
In more familiar vector calculus notation, this equation takes the form
\begin{equation}
	\eta = \text{div}(p \; \text{grad} \; \hat{\eta}).
\end{equation}
Using this isomorphism, we may now define a new metric on $\mathcal{M}$, which is essentially a probability weighted version of the Dirichlet Energy Functional. In~particular, we take:
\begin{equation} \label{Wasserstein Metric}
    G_{\mathcal{W}_2}(\eta_1, \eta_2) = \int_S p \; d\hat{\eta}_1 \wedge \star d\hat{\eta}_2 = - \int_S \eta_1 \wedge \star \hat{\eta}_2 = - \int_S \hat{\eta}_1 \wedge \star \eta_2  \, .
\end{equation}
As our notation suggests, this metric can be understood as the infinitesimal form of the Wasserstein two distance defined in Appendix \ref{Appendix Optimal Transport}. A~proof of this fact can be found in~\cite{cotler2022renormalization}. 

Now, let us compute the exterior derivative of the differential entropy, and~by extension its gradient with respect to the newly minted Wasserstein metric (\ref{Wasserstein Metric}). Let $\eta = \delta p$ denote an element of $T\mathcal{M}$ obtained by perturbing the density $p$. Moreover, let $\hat{\eta}$ denote the element isomorphic to $\eta$ via (\ref{Tangent Space Isomorphism}). We may now complete the desired computation:

\vspace{-12pt}
\begin{flalign}
    \delta S[p] &= -\int_S \left(\delta p \wedge \star \ln(p) + p \wedge \star \frac{\delta p}{p}\right) = - \int_S \eta \wedge \star(\ln(p) + 1) \\
    &= -\int_S d i_{p (d\hat{\eta})_{\musSharp}} \text{Vol}_S \wedge (\ln(p) +1) \;\;\;\;\; \text{Using (\ref{Tangent Space Isomorphism})} \\
    &= -\int_S (-1)^d i_{p (d\hat{\eta})_{\musSharp}} \text{Vol}_S \wedge \frac{dp}{p} \;\;\;\;\; \text{Integrating by parts} \\
    &= -\int_S (-1)^d p d\hat{\eta} \wedge \star \frac{dp}{p} \;\;\;\;\; i_{\omega_{\musSharp}} \text{Vol}_S = \star \omega \text{ and } \int_S \star \alpha \wedge \beta = \int_S \alpha \wedge \star \beta \\
    &= -\int_S (-1)^d d\hat{\eta} \wedge \star dp = -\int_S \hat{\eta} \wedge \star \Delta p \;\;\;\;\; \text{Integrating by parts once more} \\
    &= G_{\mathcal{W}_2}(\Delta p, \eta)   \, .
\end{flalign}
Hence, we have succeeded in showing that:
\begin{equation}
    \text{grad}_{\mathcal{W}_2} S[p] = \Delta p  \, .
\end{equation}
meaning that we can indeed write the heat equation:
\begin{equation}
    \frac{\partial}{\partial t} p_t = \text{grad}_{\mathcal{W}_2} S[p_t]  \, .
\end{equation}

In general, if~we specify a functional, $\mathcal{F}: \mathcal{M} \rightarrow \mathbb{R}$, and~follow the approach described above, the~resulting gradient flow equation will read:
\begin{equation} \label{General Continuity Equation}
\frac{\partial}{\partial t} p_t = \text{grad}_{\mathcal{W}_2} \mathcal{F}[p_t] = \star d i_{p_t d(\frac{\delta \mathcal{F}}{\delta p}\rvert_{p_t})_{\musSharp}} \text{Vol}_S \, .
\end{equation}
For our purposes, it will be interesting to consider functionals that are the sum of two~pieces:
\begin{equation}
    \mathcal{F}[p] = S[p] + \mathcal{V}[p] = -\int_S p \wedge \star \ln(p) + \int_S p \wedge \star V \, .
\end{equation}
Here, $S[p]$ is the differential entropy; thus, this part of the functional will source a diffusion equation in (\ref{General Continuity Equation}). The~second term should be interpreted as a potential and~will introduce \emph{{drift}} into (\ref{General Continuity Equation}). Provided $V: S \rightarrow \mathbb{R}$ is a function that does \emph{not} depend on $p$, (\ref{General Continuity Equation}) takes the form of a \emph{{convection--diffusion equation}}:

\begin{equation} \label{Fokker Planck main text}
    \frac{\partial}{\partial t} p_t - \star d i_{dp_{\musSharp}} \text{Vol}_S - \star d i_{p_t dV_{\musSharp}} \text{Vol}_S = 0 \, .
\end{equation}

This is precisely the \emph{{Fokker--Planck equation}}. In vector calculus notation, it takes the~form:
\begin{equation}
    \frac{\partial}{\partial t} p_t - \Delta p_t - \text{div}(p \; \text{grad} V) = 0 \, .
\end{equation}

If the normalization $Z = \int_S e^{-V} \text{Vol}_S$ is finite, $q = \frac{1}{Z} e^{-V}$ is a probability distribution, which is the stationary state of (\ref{Fokker Planck main text}). In~this case, the~gradient flow of $\mathcal{F}$ can equivalently be understood as the gradient flow of the KL-Divergence between $p_t$ and $q$. This follows from a simple calculation:
\begin{equation}
    \mathcal{F}[p_t] + \ln(Z) = \int_S p \wedge \star(\ln(\frac{p_t}{e^{-V}}) + \ln(Z)) = \int_S p_t \wedge \star \ln(\frac{p_t}{q}) = D_{KL}(p_t \parallel q) \, .
\end{equation}
Because $\ln(Z)$ is independent of the distribution $p$, the~variational derivatives of $\mathcal{F}$ and $D_{KL}(p_t \parallel q)$ will be equal: $\frac{\delta \mathcal{F}}{\delta p} = \frac{\delta D_{KL}(p \parallel q)}{\delta p}$. Thus, both of these functionals provide the same (\ref{General Continuity Equation}). 

We are now prepared to make good on our promise and show that the Wegner--Morris equation is equivalent to the Fokker--Planck equation with a specified stationary distribution. To~this end, let us write:
\begin{equation}
    p = \frac{e^{-S_p(x)}}{Z_p}, \;\;\; q = \frac{e^{-S_q(x)}}{Z_q} \, .
\end{equation}
where $Z_d = \int_S \star \hat{d}$ is the normalization factor for the Boltzmann weight of the distribution $d$, i.e.,~$\hat{p} = e^{-S_p}$. Let us also define
\begin{equation}
    \Sigma = -\ln(\frac{\hat{p}}{\hat{q}}) = S_q - S_p; \;\;\; \Psi = \text{grad}_g \Sigma \, .
\end{equation}
which are the analogs of the renormalization scheme function, and~the associated reparameterization kernel. Now, we need only compute the variation of the KL-Divergence:

\begin{flalign}
    \delta D_{KL}(p \parallel q) &= \int_S \left(\delta p \wedge \star(\ln(p) - \ln(q) + 1\right) \\
    &= \int_S \delta p \wedge \star \left(-S_p - \ln(Z_p) + 1 + S_q + \ln(Z_q)\right) \\
    &= \int_S \delta p \wedge \star \left(S_q - S_q\right) \;\;\; \int_S \delta p \wedge \star \text{const.} = 0 \text{ s.t. } \int_S \star(p + \delta p) = 1 \\
    &= \int_S di_{p d\hat{\eta}_{\musSharp}} \text{Vol}_S \wedge \Sigma \;\;\; \text{Via the tangent space isomorphism with $\delta p = \eta$} \\
    &= (-1)^d \int_S i_{p d\hat{\eta}_{\musSharp}} \text{Vol}_S \wedge d\Sigma \;\;\; \text{Integrating by parts} \\
    &= (-1)^d \int_S p d\hat{\eta} \wedge \star d\Sigma \;\;\; i_{\omega_{\musSharp}} \text{Vol}_S = \star \omega \text{ and } \int_S \star \alpha \wedge \beta = \int_S \alpha \wedge \star \beta \\
    &= - (-1)^{d} \int_S \hat{\eta} \wedge d(\star p \; \Sigma) \;\;\; \text{Integrating by parts} \\
    &= -G_{\mathcal{W}_2}(\star d(\star p \; d\Sigma)), \hat{\eta})  \, .
\end{flalign}
Hence, we have shown that:
\begin{equation}
    \text{grad}_{\mathcal{W}_2} D_{KL}(p \parallel q) = \star d(\star p d\Sigma)  \, .
\end{equation}
Thus, the~Fokker Planck equation associated with the gradient flow of $\mathcal{F}$ is precisely the Finite Dimensional Wegner--Morris equation:
\begin{equation} \label{Finite Dimensional Wegner--Morris}
    \frac{\partial}{\partial t} p_t = -\text{grad}_{\mathcal{W}_2} D_{KL}(p \parallel q) = -\star d(\star p \; d\Sigma) = \frac{\partial}{\partial x^i}\left(p g^{ij} \frac{\partial}{\partial x^j} \Sigma\right) = \frac{\partial}{\partial x^i}\left(\psi^i \; p\right)  \, .
\end{equation}

\subsection{Renormalization Group Flow as Optimal~Transport}

Relating RG flow to the Continuity Equation~(\ref{Finite Dimensional Wegner--Morris}) is more or less an exercise in properly identifying the sample space of the RG probability distribution and~completing the analogies that follow from there. For~simplicity, we have reproduced this exercise in the form of a dictionary between optimal transport for finite sample spaces and ERG which can be found in Table \ref{OTRG}.

\begin{landscape}

\begin{table}[]
\centering
\begin{tabular}{lll}
                                   & \textbf{Optimal Transport}                                                                         & \textbf{Exact Renormalization Group}                                                                                                                                                                                                          \\
\textbf{Sample Space}              & $S$, a differentiable manifold                                                                     & $\mathfrak{F}(M)$, the space of fields on spacetime $M$                                                                                                                                                                                       \\
\textbf{Coordinates}               & $y \in S$, a point                                                                                 & $\phi \in \mathfrak{F}(M)$, a field                                                                                                                                                                                                           \\
\textbf{Indices}                   & $y^i, \; i \in \{1, ..., \text{dim}(S)\}$, components                                              & $\phi(x), \; x \in M$, image at each point in $M$                                                                                                                                                                                             \\
\textbf{Metric}                    & $g = g_{ij} dy^i \otimes dy^j$                                                                     & $\dot{C}_{\Lambda}(x,y)$, an integral kernel                                                                                                                                                                                                  \\
\textbf{Space of Distributions}    & $\mathcal{M} := \{p: S \rightarrow \mathbb{R} \; | \; \int_S \; p \; Vol_S = 1\}$                  & $\mathcal{M} := \{P: \mathfrak{F}(M) \rightarrow \mathbb{R} \; | \; \int_{\mathfrak{F}(M)} \mathcal{D} \phi \; P[\phi] = 1\}$                                                                                                                 \\
\textbf{Tangent Space}             & $T\mathcal{M} := \{\eta: S \rightarrow \mathbb{R} \; | \; \int_S \; \eta \; \text{Vol}_S = 0\}$    & $T \mathcal{M} := \{\eta: \mathfrak{F}(M) \rightarrow \mathbb{R} \; | \; \int_{\mathfrak{F}(M)} \; \mathcal{D} \phi \; \eta[\phi] = 0\}$                                                                                                      \\
\textbf{Tangent Space Isomorphism} & $\eta = \star d i_{p d\hat{\eta}_{\musSharp}} \text{Vol}_S$                                        & $\eta = \int_{M \times M} \text{Vol}_M(x) \wedge \text{Vol}_M(y) \; \frac{\delta}{\delta \phi(x)} \left(P \dot{C}_{\Lambda}(x,y) \frac{\delta}{\delta \phi(y)} \hat{\eta}[\phi] \right)$                                                      \\
\textbf{Wasserstein Metric}        & $G_{\mathcal{W}_2}(\eta_1, \eta_2) = \int_S p \; d\hat{\eta}_1 \wedge \star d\hat{\eta}_2$         & $\mathcal{G}_{\mathcal{W}_2}(\eta_1, \eta_2) = \int_{\mathfrak{F}(M)} \mathcal{D} \phi P[\phi] \; \int_{M \times M} \dot{C}_{\Lambda}(x,y) \frac{\delta \hat{\eta}_1[\phi]}{\delta \phi(x)} \frac{\delta \hat{\eta}_2[\phi]}{\delta \phi(y)}$ \\
\textbf{ERG Kernel}                & $\Sigma(x) = -\ln(\frac{\hat{p}}{\hat{q}})$                                                        & $\Sigma[\phi] = -S[\phi] + 2 \Tilde{S}[\phi]$                                                                                                                                                                                                 \\
\textbf{Reparameterization Kernel} & $\Psi(x,p) = \text{grad}_g \Sigma(x)$                                                              & $\Psi[\phi,x] = -\frac{1}{2} \int_M \text{Vol}_M(y) \dot{C}_{\Lambda}(x,y) \frac{\delta \Sigma[\phi]}{\delta \phi(y)} $                                                                                                                       \\
\textbf{Potential Function}        & $V: S \rightarrow \mathbb{R}$, Stationary Distribution & $2\Tilde{S}[\phi]$, Twice the Seed Action                                                                                                                                                                                   \\
\textbf{Wegner-Morris Equation}    & $\text{grad}_{\mathcal{W}_2} D_{KL}(p_t \parallel q_t) = \star d(\star p d\Sigma)$                 & $\text{grad}_{\mathcal{W}_2} D_{KL}(P_{\Lambda} \parallel Q_{\Lambda}) = \int_M \text{Vol}_M(x)  \frac{\delta}{\delta \phi(x)} \left(\Psi_{\Lambda}[\phi,x] P_{\Lambda}[\phi]\right)$                                                        
\end{tabular}
\caption{A dictionary describing the correspondence between Optimal Transport and ERG.}
\label{OTRG}
\end{table}

\end{landscape}

\section{From Stochastic Differential Equations to Partial Differential Equations and Back~Again} \label{SDEs and PDEs}

Having placed the Exact Renormalization Group flow equations squarely in the context of Partial Differential Equations, we would now like to take a brief detour into addressing some of the generic properties of these equations. In~particular, we will be interested in understanding the relationship between continuity equations on the one hand and~Stochastic Differential Equations on the other. For~more information on the relationship between Stochastic Differential Equations, partial differential equations, and~optimal transport, see~\cite{bogachev2022fokker,da2004kolmogorov,fuhrman2002nonlinear,chengeometric,shreve2004stochastic,ambrosio2005gradient,santambrogio2017euclidean,villani2009optimal}.

\subsection{Stochastic Differential~Equations}

A stochastic process is a one-parameter family of random variables, $\{X_t\}_{t\geq 0}$. For~our purposes, we shall be interested in stochastic processes whose dynamics are governed by a Stochastic Differential Equation of the Ito form ~\mbox{\cite{ito1944109,ito1951stochastic,ito1996diffusion}}:
\begin{equation} \label{SDE Drift Diffusion}
    dX_t^i = m^i(X_t, t) dt + \sigma^i_j(X_t,t) dW_t^j   \, .
\end{equation}
Here, $m^i$ describes the \emph{{drift}}, or~mean rate of change, while $\sigma^i_j$ describes the \emph{{diffusion}}, or~mean variability. $dW_t^i$ corresponds to an independent increment of a Brownian motion, a~random variable drawn from a standard normal distribution corresponding to the noisy motion of the process over a single increment of~time. (We recall that {the Ito formulation is one of two major approaches to the subject of stochastic calculus, with the other being the Stratonovich formalism. Here, we use Ito formulation because it makes the relationship between Stochastic Differential Equations and Partial Differential Equations very clear}).

Physicists may find it useful to compare (\ref{SDE Drift Diffusion}) to the \emph{{Langevin Equation}}
\begin{equation}
    \frac{d X^i}{dt} = \mu^i(X_t) + \eta^i_t  \, .
\end{equation}
which describes the dynamics of a quantity $X_t$ governed by a law $\frac{dX}{dt} = \mu$ but~subject to random fluctuations, $\eta$ \cite{coffey2012langevin,sekimoto1998langevin}. To~match the Stochastic Differential Equation, we should take $\eta_t = (\eta^1_t, ..., \eta^n_t)$ to be a random variable with the correlation structure
\begin{equation}
    \langle \eta^i_t \eta^j_{t'} \rangle = \delta^{kl} \sigma^i_k \sigma^j_l \delta(t - t')  \, .
\end{equation}

Given a function: $f: [0,T] \times S \rightarrow \mathbb{R}$, we can determine its stochastic differential when evaluated over the process $X_t$ by using the principal of quadratic variation. Quadratic variation dictates that the product of two increments of Brownian motion scales linearly in the interval between realizations of the stochastic process. This fact follows from the observation that the variance of a Weiner process is linear in time. Schematically, we encode the principal of quadratic variation in the form:
\begin{equation}
    dW_t^i dW_t^j \sim \delta^{ij} dt + (1-\delta^{ij}) O(dt^{2})  \, .
\end{equation}
If we now expand the differential of $f(X_t, t)$ in a power series, retaining terms up to $O(dt)$, we find:
\begin{equation} \label{SDE of function}
    df(X_t,t) = \left(\frac{\partial f}{\partial t} + m^i \frac{\partial f}{\partial x^i} + \frac{1}{2} \delta^{kl} \sigma^{i}_{k} \sigma^{j}_{l} \frac{\partial^2 f}{\partial x^i \partial x^j}\right) dt + \frac{\partial f}{\partial x^i} dW_t^i   \, .
\end{equation}
The second-order derivative terms are the unique addition provided to us by quadratic variation. The~realizations, $\{f(X_t,t)\}_{t \geq 0}$, now define a stochastic process in their own right, governed by the Stochastic Differential Equation~(\ref{SDE of function}). 

A stochastic process, $\{Y_t\}_{t \geq 0}$, is called a \emph{{Martingale}} if it satisfies the equation
\begin{equation} \label{Martingale Definition}
    \mathbb{E}(Y_t \; | \{Y_s\}_{s \leq \tau}) = Y_{\tau}; \;\;\; \forall s \leq t   \, .
\end{equation}
One can read (\ref{Martingale Definition}) as specifying that the mean value of the stochastic process, $Y_t$, has no tendency to change over time. Inspired by this interpretation, it is not hard to show that a stochastic process will be a Martingale if and only if it is described by a Stochastic Differential Equation with vanishing drift. This leads to a beautiful connection between the theory of stochastic processes, and~harmonic analysis~\cite{simon2015harmonic,maruyama1949harmonic,bochner2005harmonic}. Considering the stochastic process $\{f(X_t,t)\}_{t \geq 0}$, if~we demand that the drift in (\ref{SDE of function}) be set equal to zero, we find that $f$ must satisfy the Partial Differential Equation:
\begin{equation} \label{Backward Equation}
    \frac{\partial f}{\partial t} + m^i \frac{\partial f}{\partial x^i} + \frac{1}{2} \delta^{kl} \sigma^{i}_{k} \sigma^{j}_{l} \frac{\partial^2 f}{\partial x^i \partial x^j} = 0 \, .
\end{equation}
Moreover, using (\ref{Martingale Definition}), we see that this partial differential equation has a formal solution:
\begin{equation} \label{Solution to backwards equation}
    f(X_t,t) = \mathbb{E}(f(X_T,T) \mid \{X_s\}_{s\leq t})   \, .
\end{equation}
where $T$ should be regarded as the terminal time of the stochastic~process. 

Let us define the differential operator that appears in (\ref{Backward Equation}) as:
\begin{equation} \label{Generator of Diffusion}
    \mathcal{A} = m^i \frac{\partial}{\partial x^i} + \frac{1}{2} \delta^{kl} \sigma^{i}_{k} \sigma^{j}_{l} \frac{\partial^2}{\partial x^i \partial x^j} \, .
\end{equation}
With respect to the $\ell^2$ inner product on the space of functions, we can define a formal adjoint operator: $\mathcal{A}^{\dagger}$ such that
\begin{equation}
    G_{\ell^2}(f, \mathcal{A}(g)) = G_{\ell^2}(\mathcal{A}^{\dagger}(f), g) \, .
\end{equation}
The adjoint operator can be deduced by integrating by parts, and~we find:
\begin{equation}
    \mathcal{A}^{\dagger}(f) = -\frac{\partial}{\partial x^i}(m^i f) + \frac{1}{2}\frac{\partial^2}{\partial x^i \partial x^j}( \delta^{kl} \sigma^{i}_{k} \sigma^{j}_{l} f) \, .
\end{equation}

The differential equation generated by the adjoint operator therefore reads:
\begin{equation} 
    \frac{\partial f}{\partial t} = \frac{\partial}{\partial x^i}(m^i f) - \frac{1}{2}\frac{\partial^2}{\partial x^i \partial x^j}( \delta^{kl} \sigma^{i}_{k} \sigma^{j}_{l} f) \, .
\end{equation}
This is the \emph{{Fokker--Planck}} equation. In fact, it is slightly more general than the Fokker--Planck equation we uncovered through our analysis in Section~\ref{WM = FP} because it allows for a stochastic process with non-trivial diffusion matrices $\sigma^i_j$. To~reconcile the Fokker--Planck Equation~(\ref{Fokker Planck main text}), we should consider the Stochastic Differential Equation:
\begin{equation} \label{Fokker--Planck SDE}
    dX_t^i = -(\text{grad}_g V)^i dt + \sqrt{2} \delta^{i}_j dW_t^j \, .
\end{equation}
Here, we again see the role of the potential $V$ in sourcing the mean drift behavior of the stochastic process. That is, the~stochastic behavior associated with the random variable described by a probability distribution solving the Fokker--Planck equation with potential function $V$ is itself a stochastic gradient flow with respect to that~potential. 

The pair of differential equations we have written have interpretations as ``Forward'' and ``Backward'' processes. The~equation generated by $\mathcal{A}$:
\begin{equation}
    \left(\frac{\partial}{\partial t} + \mathcal{A}\right)f = 0  \, ,
\end{equation}
is called the \emph{{Backward}} equation because its solution, i.e.,~(\ref{Solution to backwards equation}), is specified by a terminal condition. On~the contrary, the~adjoint equation:
\begin{equation}
    \left(\frac{\partial}{\partial t} - \mathcal{A}^{\dagger}\right) f = 0 \, ,
\end{equation}
is called the \emph{{Forward}} equation because its solution is specified by an initial condition: $f(0,x)$.

\subsection{Continuity~Equations}

Let us now begin from the reverse perspective and~seek to understand how a continuity equation might be associated with a stochastic process. A~\emph{{measure}} on $S$ is a top form whose integral over all of $S$ can be normalized to $1$. For~convenience, we will refer to such a form as $\mu \in M_{S}$ with
\begin{equation}
    M_{S} := \{\mu \in \Omega^{d}(S) | \; \int_{S} \mu = 1\} \, .
\end{equation}
By Hodge duality, a~measure can be related to a distribution: $p = \star \mu$, or~$\mu = p \text{Vol}_{S}$, which would bring us back into the notation used in the previous section. We will move back and forth between the measure-based and~probability-density-based approaches more or less at~will. 

To write down a continuity equation, we consider a one-parameter family of measures, or~otherwise a trajectory through the space $M_{S}$, $\mu: [0,1] \rightarrow M_{S}$, which we denote as $\mu_t(x) \in M_{S}$, governed by the differential equation:
\begin{equation} \label{Continuity}
    \frac{\partial}{\partial t} \mu_t + di_{v_t} \mu_t = 0  \, .
\end{equation}
Here, $v_t: [0,1] \rightarrow TS$ is a time-dependent vector field. Since $\mu_t$ is a top form, we can regard the second term above as the Lie Derivative and write:
\begin{equation}
    \frac{\partial}{\partial t} \mu_t + \mathcal{L}_{v_t} \mu_t = 0 \, .
\end{equation}
Thus, a~continuity equation has the immediate interpretation as a flow generated by the vector field $v_t$ \cite{ambrosio2005gradient}. Compared to (\ref{Fokker Planck main text}), we see that the vector field that generates a gradient flow with respect to the potential $\mathcal{F}$ is given by:
\begin{equation}
    v_t = -d\left(\frac{\delta \mathcal{F}}{\delta p}(p_t)\right)_{\musSharp} \, .
\end{equation}

A measure $\mu_t$ that solves the continuity equation is referred to as a \emph{strong solution}. However, we can also consider a weaker condition in which the integral of (\ref{Continuity}) against a compactly supported $C^1(S)$ function is always zero~\cite{santambrogio2015optimal}. That is, given $\psi: [0,1] \rightarrow C^1(S)$, we demand:
\begin{equation} \label{Weak Solution}
    0 = \int_{[0,1] \times S} \psi_t (\frac{\partial}{\partial t} \mu_t + di_{v_t} \mu_t) \, .
\end{equation}
Integrating by parts and using the fact that $\psi_t$ is compactly supported (\ref{Weak Solution}) implies that
\begin{equation} \label{Weak sol 2}
    0 = -\int_{[0,1]\times S} \mu_t (\frac{\partial \psi_t}{\partial t} + d\psi_t(v_t)) \, .
\end{equation}
To arrive at \eqref{Weak sol 2} we have 
also used the fact that
\begin{equation}
    d\psi_t \wedge i_{v_t} \mu_t = d\psi_t(v_t) \mu_t \, ,
\end{equation}
where $d\psi_t(v_t)$ is the pairing of $d\psi_t$ and $v_t$ in the sense of forms and tangent vectors.
We can interpret \eqref{Weak sol 2} as the condition:
\begin{equation}
    \int_{[0,1]} \mathbb{E}_{\mu_t}(\frac{\partial \psi_t}{\partial t} + d\psi_t(v_t)) = 0  \, .
\end{equation}

Provided we can exchange the order of the t-derivative and the integral over $S$, we can reframe our analysis in terms of a $t$-independent function, $\phi: S \rightarrow \mathbb{R}$, in~which case we find
\begin{equation} \label{Ehrenfest}
    \frac{d}{dt}\mathbb{E}_{\mu_t}(\phi) + \mathbb{E}_{\mu_t}(d\phi(v_t)) = 0  \, .
\end{equation}
This equation can be regarded as an Ehrenfest theorem specifying the mean dynamics of the field $\phi$. In~particular, it says that $\phi$ will satisfy a gradient flow in the direction of $v_t$, and~thus we can regard (\ref{Ehrenfest}) as the expectation value of the Stochastic Differential Equation~(\ref{SDE Drift Diffusion}). 

The heat equation is a special case of the continuity equation with the flow vector given by $\frac{\text{grad}p_t}{p_t} = \text{grad}(\ln(p_t))$: to see that this is the case, let us write $\mu_t = p_t \text{Vol}$. Then, it is straightforward to show that
\begin{equation}
    di_{\frac{\text{grad}p_t}{p_t}}\mu_t = di_{\text{grad}p_t} \frac{p_t \text{Vol}_{S}}{p_t} = di_{\text{grad}p_t} \text{Vol}_{S} = \text{div}(\text{grad}p_t) \text{Vol}_{S} = \Delta p_t \text{Vol}_{S} = \Delta \mu_t \, .
\end{equation}
Here we have used that the Laplacian operator is defined on differential forms as $\Delta = d\star d \star + \star d \star d$. It is easy to show that the Laplacian of a top form, $\mu_t = p_t \text{Vol}$, is related to the Laplacian of its associated scalar by:
\begin{equation}
    \Delta \mu_t = (d \star d \star + \star d \star d)\mu_t = d \star d \star \mu_t = d \star d p_t = \star \Delta p_t \, .
\end{equation}
Hence, in this case (\ref{Continuity}) becomes:
\begin{equation}
    \frac{\partial}{\partial t} \mu_t + \Delta \mu_t = 0; \;\;\; \frac{\partial}{\partial t} p_t + \Delta p_t = 0 \, .
\end{equation}

The fundamental solution to the heat equation with initial data $p_0$ is written formally in terms of the heat kernel~\cite{davies1989heat,berline2003heat}:
\begin{equation}
    p_t = e^{-t\Delta} p_0  \, ,
\end{equation}
with $e^{-t\Delta}$ the operator obtained by exponentiating $\Delta$, which acts on the initial data. {Indeed, it is easy to show that:
\begin{equation}
    \frac{\partial}{\partial t}(e^{-t\Delta} p_0) = -\Delta e^{-t\Delta} p_0 \, .
\end{equation}
}
In a position basis, we can represent the heat kernel as an integral kernel of the form:
\begin{equation}
    \bra{x} e^{-t\Delta} \ket{y} = H(x,y,t) = (4\pi t)^{-d/2} e^{-\frac{\norm{x-y}^2}{4t}}  \, .
\end{equation}
Thus, we can interpret the heat kernel as the transition function for a Markov process corresponding to the probability density of translation from sample point $x$ to sample point $y$ in an interval $2t$. This density of the form of a multivariate Gaussian, which we shall denote by $H(x,y,t) = \mathcal{N}(y, 2t\mathbb{I})$ ($\mathcal{N}(\mu, \Sigma)$, is the multivariate normal distribution with mean parameter $\mu$ and covariance matrix $\Sigma$).

A general continuity equation will be generated by a differential operator $\mathcal{A}$, through the equation:
\begin{equation} \label{A Continuity}
    \left(\frac{\partial}{\partial t} + \mathcal{A}\right) p_t = 0  \, .
\end{equation}
We can play the same game as before and solve this differential equation formally by writing $p_t = e^{-t\mathcal{A}} p_0$. Again, we can choose a continuous basis and express the heat kernel of the operator $\mathcal{A}$ as an integral kernel:
\begin{equation}
    \bra{x} e^{-t\mathcal{A}} \ket{y} = \mathcal{H}(x,y,t)  \, .
\end{equation}
It remains natural to interpret $\mathcal{H}(x,y;t)$ as the transition probability density for a Markov process~\cite{rudnicki2002markov,lorenzi2006analytical}. The~Markov process in question is defined by a continuous set of operators, $\{P_t\}_{t \geq 0}$. One can regard the operator $P_t$ as the ``time evolution operator'' associated with $\mathcal{A}$, which, in~this context, should be understood as the infinitesimal generator of time translations. In~other words, $P_t$ is obtained by exponentiating the operator $\mathcal{A}$, $P_t = e^{-t\mathcal{A}}$. 

Given any measurable function, $f: S \rightarrow \mathbb{R}$, the~action of the operator $P_t$ translates $f$ along the Markov chain. If~we work in the positional basis, we can write:
\begin{equation} \label{Markov Operator}
    P_t(f) = e^{-t\mathcal{A}}(f) = \int_{S} \text{Vol}_{S}(z) \mathcal{H}(z,x,t) f(z)  \, .
\end{equation}
which is simply the convolution of the transition density $\mathcal{H}$ with $f$. This provides us with a very useful formula for determining the operator $\mathcal{A}$:
\begin{equation} \label{Markov Process to PDE}
    \mathcal{A}(f) = \lim_{t \rightarrow 0} \frac{P_t(f) - f}{t}  \, .
\end{equation}
Equation~(\ref{Markov Process to PDE}) reverses the operator exponentiation by differentiating along the integral curve $P_t(f)$.

The family of operators $\{P_t\}_{t \geq 0}$ can be regarded as a \emph{{semigroup}} with the simple composition law~\cite{kolokoltsov2011markov}:
\begin{equation}
    P_{t+s}(f) = P_t(P_s(f)) = P_s(P_t(f))  \, .
\end{equation}
Moreover, the~set of operators $\{P_t\}_{t \geq 0}$ defines a stochastic process $\{X_t\}_{t\geq 0}$, satisfying the law~\cite{chengeometric}:
\begin{equation}
    \mathbb{E}(f(X_{s+t})\mid X_{r \leq t}) = P_s(f)(X_t)  \, .
\end{equation}
This expression is equivalent to the Martingale condition (\ref{Solution to backwards equation}). Writing the operator $\mathcal{A}$ in the form (\ref{Generator of Diffusion}), the~stochastic process $\{X_t\}_{t \geq 0}$ is immediately identified with the Stochastic Process:
\begin{equation}
    dX_t^i = m^i(X_t,t) dt + \sigma^i_j dW_t^i  \, .
\end{equation}
Thus, we have succeeded in mapping our way back to a Stochastic Differential Equation, this time beginning from a diffusion~equation.

\section{Dynamical Bayesian Inference and the Backward~Equation} \label{Dynamical Bayes}

The formulation of the Exact Renormalization Group flow as an Optimal Transport problem, particularly in its relationship to the extremization of relative entropy, already suggests a deep connection between renormalization, diffusion, and~information theory. We now turn our attention to the task of fleshing out this relationship and, in~doing so, providing an information-theoretic conceptualization of renormalization through the intermediary of statistical inference. To~accomplish this task, we will build on the language of \emph{{Dynamical Bayesian Inference}} introduced in~\cite{berman2022dynamics}. We will show that the dynamics of an inferred probability model defined by a continuously updating Bayesian scheme give rise to an \emph{inverse} process governed by a diffusion equation that can be brought into correspondence with ERG. Using this fact, we advocate for the perspective that renormalization can be understood as the {\it{{inverse}}} process of statistical~inference. 

We should emphasize that inversion is understood in the statistical sense as a `reverse time' stochastic process relative to the Kolmogorov forward process defined by ERG~\cite{anderson1982}. Such stochastic processes have recently appeared in diffusion learning, where the reverse time Stochastic Differential Equation relative to a forward diffusion process is modeled via a score-based generative algorithm~\cite{Song:2020hus}. From~the point of view developed in this section, a~score-based generative algorithm can be interpreted as implementing a form of Bayesian inference. We will revisit this and other related points in Section~\ref{Discussion}.

\subsection{Bayesian~Inversion}

Bayesian inference is a probabilistic, self-consistent approach to adjusting one's beliefs in the presence of new information. It originates from Bayes' Law, which encodes the probabilistic relationship between two, potentially co-varying, random variables $H \in S_H$ and $E \in S_E$:
\begin{equation} \label{Bayes}
    P(H \mid E) = \frac{P(E \mid H) P(H)}{P(E)}  \, .
\end{equation}
In Bayesian Inference, the~variables $H$ and $E$ are respectively identified with the \emph{{Hypothesis}} and the \emph{{Evidence}}. In~this language, one can attach a compelling geometric interpretation to Bayes' law as measuring the relative volumes of the region $\mathcal{H} \subset S_H \times S_E$, for which both $H$ and $E$ are realized, and~the region $\mathcal{O} \subset S_E$, for which the observed evidence is true. Thus, the~interpretation of Bayesian Inference is that it measures the volume of hypotheses that is consistent with a given set of observed data
, which we denote as the region $\mathcal{H}\mid \mathcal{O} \subset S_H \times S_E$.   

A particularly interesting form of Bayesian Inference arises in the context of so-called Bayesian Inversion problems~\cite{dashti2017bayesian, hoang2013complexity,cockayne2016probabilistic,raissi2019physics,adler2018deep,schillings2016scaling,matthies2016parameter}. There, we are concerned with a pair of random variables, $Y \in S_Y$ and $U \in S_U$, which are related by the following process:
\begin{equation} \label{Channel Process}
    y = G(u) + n
\end{equation}
We regard $G: S_U \rightarrow S_Y$ as a deterministic function, denoted by $N \in S_N \simeq S_Y$ noise, which can either be associated with the process itself or~the process of measuring its~output
. 

The goal of the Bayesian Inversion problem is to determine the signal, $u$, that led to the measured output, $y$. However, it is often impractical to truly invert the process; thus, it is more prudent to seek to place a probability distribution over the~inputs.

To this end, we regard $U \in S_U$ as having a prior measure $\Pi_0 = \pi_0(u) \text{Vol}_U(u) \in M_{S_U}$ and~treat the noise $N$ as a random variable independent from $Y$ also possessing of a measure $\rho_0 = p_0(n) \text{Vol}_N(n) \in M_{S_N}$. Then, we can define the conditional random variable $Y \mid U$ as:
\begin{equation}
    Y \mid U = G(U) + N \, .
\end{equation}

If we regard $\phi: S_Y \rightarrow S_N$ as a map from output to noise, such that
\begin{equation}
    n = \phi(y) = y - G(u)   \, .
\end{equation}
we can obtain a conditional measure for $Y \mid U$ by pulling back the measure $\rho_0$ via the map~$\phi$:

\begin{equation}
    \rho_{Y \mid U}(y \mid u) = \phi^*(\rho_0(n)) = p_0(y - G(u)) \det\left(\frac{\partial \phi^i}{\partial y^j}\right) \text{Vol}_N(y - G(u)) = p_{0}(y - G(u)) \text{Vol}_Y(y) \, .
\end{equation}
We regard the pulled-back density $p_0(y - G(u))$ as the \emph{{likelihood function}} for an inference problem and~often denote it by $p_{Y \mid U}(y \mid u)$. We therefore obtain a joint measure:
\begin{equation}
    \mu_{Y,U}(y,u) = \rho_{Y \mid U}(y \mid u) \wedge \Pi_0(u) = p_0(y - G(u)) \pi_0(u) \text{Vol}_Y(y) \wedge \text{Vol}_U(u) 
\end{equation}
for the random variable $(Y,U) \in S_Y \times S_U$.

Provided the marginalization:
\begin{equation}
    p_Y(y) = \int_{S_U} p_0(y - G(u)) \pi_0(u) \text{Vol}_U(u)
\end{equation}
is greater than zero and finite, we can define the Bayesian posterior measure as:
\begin{equation}
    \Pi_{y_*}(u) = \frac{1}{p_Y(y_*)} \rho_{Y \mid U}(y_* \mid u) \wedge \Pi_0(u) \, ,
\end{equation}
which is just a recapitulation of Bayes' law as written in the form (\ref{Bayes}). The~notation $\Pi_{y_*}(u)$ is meant to remind us that the posterior distribution depends on the realization of observed data $y_* \in S_Y$. We can also recognize $p_Y(y)$ as the marginal density for the random variable $Y$ modulo the prior distribution $\pi_0$. 

Next, we define the \emph{{Bayesian Potential}} as:
\begin{equation}
    \Phi(u; y) := -\ln(p_0(y - G(u)))
\end{equation}
which is nothing but the negative log likelihood. With~this potential in hand, we can write the Bayesian Inference condition suggestively as:
\begin{equation}
    \frac{d\Pi_{y}}{d\Pi_0}(u) = \frac{1}{p_Y(y)} e^{-\Phi(u;y)} \, .
\end{equation}
On the left hand side, we have the \emph{{Radon--Nikodym}} derivative of the posterior measure with respect to the prior measure. On~the right hand side, we have what we would like to interpret as the stationary distribution modulo observations $y$. This equation can be interpreted as follows: given any measurable function $f: S_U \rightarrow \mathbb{R}$
\begin{equation}
    \mathbb{E}_{\Pi_y}\left(f(U)\right) = \mathbb{E}_{\Pi_0}\left(\frac{d\Pi_y}{d\Pi_0}(U) f(U)\right)
\end{equation}
meaning that the expectation value with respect to the posterior measure is the same as the expectation value with respect to the prior provided the function is augmented by the evidence in the form of the stationary distribution $\frac{e^{-\Phi(u;y)}}{p_Y(y)}$. 

\subsection{Dynamical~Bayes}

\emph{{Dynamical}} Bayesian Inference is an extension of the conventional approach to Bayesian Inference in which one implements Bayesian inference as an iterative process where swathes of evidence are collected, and~the posterior distribution at the end of a given iteration is used as the prior distribution in the following iteration~\cite{berman2022dynamics}. 

Using such an approach, one can regard Bayesian Inference as a dynamical system governed by a first-order differential equation. To~begin, we define a timelike variable, $T$, which essentially corresponds to the total number of data point observed. Inference up to the ``time'' $T$ therefore leverages evidence from the set of observed data $\{y_t\}_{0 \leq t \leq T}$, which we can regard as a continuous time stochastic process coming from the data-generating measure $\mu^*_Y(y) = p^*_Y(y) \text{Vol}_Y$. In what follows {we shall regard the data-generating measure as belonging to the parametric family $p_{Y \mid U}$, associated with the true underlying signal value $u^*$. That is, $p^*_Y(y) = p_{Y \mid U}(y \mid u^*)$}. The equation satisfied by $\pi_T(u)$ is then given by:
\begin{equation} \label{Dynamical Bayes Equation}
    \frac{\partial}{\partial T} \pi_T(u) = -\left(D(p_{Y\mid U}) - \mathbb{E}_{\Pi_T}(D(p_{Y \mid U}(y \mid U)) \right) \pi_T(u)  \, ,
\end{equation}
where
\begin{equation}
    D(p) = D_{KL}(p^*_Y \parallel p)
\end{equation}
is the KL-Divergence between a distribution $p$ on $S_Y$ and the data-generating distribution~$p^*_Y$. {As an aside, we note that \eqref{Dynamical Bayes Equation} is equivalent to the replicator dynamic which appears in evolutionary game theory provided the fitness of a particular model $p_{Y\mid U}(y \mid u)$ is taken to be minus its KL-Divergence with the data-generating model. We refer the reader to~\cite{harper2009information,harper2009replicator} for more detail}.

Among the most significant insights of this approach is that, if~we consider models that are within an $\epsilon$ neighborhood of the true underlying signal, the~$2n$-point functions described by $\Pi_T$ are very well approximated by power laws:
\begin{equation} \label{Late T behavior}
    C^{i_1 ... i_{2n}} := \mathbb{E}_{\Pi_T}\left(\prod_{j = 1}^{2n} (U - u_*)^{i_j} \right) = \frac{1}{T^n} \sum_{p \in \mathcal{P}_{2n}^2} \prod_{(r,s) \in p} \mathcal{I}^{i_r i_s}\bigg\rvert_{u_*}  \, .
\end{equation}
Here, $\mathcal{I}^{ij}$ is the inverse of the Fisher Metric arising from the family of distributions $p_{Y \mid U}$, evaluated at the data-generating parameter $u_*$. 

\subsection{Bayesian Diffusion for Normal~Data}

To illustrate the Dynamical Bayesian Inference dynamic, it will be beneficial to work through the problem of performing Bayesian inference on the mean of normally distributed data with known variance, $\sigma^2$. In~this case, we take:
\begin{equation}
    y = G(u) + n
\end{equation}
where $n$ is noise distributed according to a distribution $\mathcal{N}(0,\sigma^2)$:
\begin{equation}
    p_0(n) = \frac{1}{\sqrt{2\pi \sigma^2}} e^{-\frac{1}{2\sigma^2} n^2}
\end{equation}
and $G(u) = \mu$ is a random draw from the distribution over means for the data $y$ so that $p_0(y - \mu) = \mathcal{N}(\mu,\sigma^2)$ ({for notational simplicity, we shall denote this distribution by $p(y \mid \mu)$}). The data-generating distribution is taken to belong to the same parametric family of distributions, only with a fixed but unknown ``true'' underlying mean parameter \mbox{$\mu_*$ -- $p^*_Y(y) = p(y \mid \mu_*)$}. 

The governing equation of Dynamical Bayesian Inference can be solved formally as:
\begin{equation}
    \pi_T(u) = \exp\left(-T \; D(u)\right) \exp\left(\int_{T_0}^T dT' \; D(\alpha_{T'})\right)  \, .
\end{equation}
Here, we have used an abbreviated notation in which $D(u) = D_{KL}(u_*\parallel u)$ is the KL-divergence between two distributions of the same paramteric form with given parameter values $u$. Notice that only the first exponential depends on the variable $u$; hence, we conclude that the role of the second exponential is simply to maintain the normalization of $\pi_T$ as a probability distribution. Thus, we can write:
\begin{equation}
    \pi_T(u) = \frac{1}{Z} \exp\left(-T D(u)\right)
\end{equation}
where
\begin{equation}
    Z = \int \text{Vol}_U \; \exp\left(-T \; D(u)\right) = \exp\left(-\int_{T_0}^T dT' \; D(\alpha_{T'})\right)  \, .
\end{equation}

In the case of the normal model with fixed variance, the~KL-divergence is given by $D(\mu) = \frac{1}{2\sigma^2}(\mu - \mu_*)^2$. The~$T$-Posterior is therefore given by:
\begin{equation}
    \pi_T(\mu) = \frac{1}{Z} e^{-\frac{T}{2 \sigma^2} (\mu - \mu_*)^2}  \, .
\end{equation}
It is straightforward to determine the normalization of this distribution by performing the requisite integral that is now Gaussian. When all is said and done, the~$T$-dependent posterior density is of the form:
\begin{equation} \label{Brownian Posterior}
    \pi_T(\mu) = \frac{1}{\sqrt{2\pi (\sigma^2/T)}} e^{-\frac{1}{2 (\sigma^2/T)}(\mu - \mu_*)^2} \, .
\end{equation}

Let us now compare (\ref{Brownian Posterior}) to the standard density of a length $t$ increment of a Weiner process with diffusivity parameter $\sigma$: 
\begin{equation}
    f_{W_t}(x) = \frac{1}{\sqrt{2\pi \sigma^2 t}} e^{- \frac{x^2}{2\sigma^2 t}}  \, .
\end{equation}
Recall that the density $f_{W_t}(x)$ solves the heat equation:
\begin{equation}
	\frac{\partial}{\partial t} f_{W_t}(x) - \sigma^2 \frac{\partial^2}{\partial x^2} f_{W_t}(x) = 0.
\end{equation}
Following the analysis of Section~\ref{SDEs and PDEs}, one can also recognize the stochastic process $\{f_{W_t}(X_t)\}_{t \geq 0}$ as a martingale adapted to the Weiner process
\begin{equation}
	dX_t = \sigma dW_t.
\end{equation}
We therefore recognize (\ref{Brownian Posterior}) as describing a shifted Brownian motion for the mean parameter with diffusivity $\sigma$ in the ``time'' parameter $\tau = \frac{1}{T}$:
\begin{equation}
    \pi_{\tau}(\mu) = \frac{1}{\sqrt{2\pi \sigma^2 \tau}} e^{-\frac{(\mu - \mu_*)^2}{2\sigma^2 \tau}} \, .
\end{equation}

This observation lends credence to the idea that Bayesian inference can be associated with a diffusion process. Moreover, it provides an important insight: Bayesian \emph{{Diffusion}} ensues \emph{{backwards}} with respect to the performance of Bayesian \emph{{Inference}}, in~the timelike parameter, $\tau$, which is the \emph{{inverse}} of the Bayesian time, $T$, originally~introduced. 

Given the $\tau$-posterior $\pi_{\tau}$, and~terminal data such as the parametric form of the data-generating distribution, one can obtain the $\tau$ path of the posterior predictive distribution for future data by marginalizing. For~the normal model, this means:
\begin{equation}
    p_{\tau}(y) = \int_{\mathbb{R}} d\mu \; \pi_{\tau}(\mu) p(y \mid \mu) = \int_{\mathbb{R}} d\mu \; \pi_{\tau}(\mu) p_{0}(y - \mu)  \, ,
\end{equation}
which we can recognize as the convolution of $\pi_{\tau}$ and $p_0(y)$ and~interpret as the action of the Green Function of the Heat Operator translating the model forward in $\tau$ (and backward in $T$). 

\subsection{Bayesian Drift and Scheme~Independence} 

We have now shown that the solution to the Dynamical Bayesian inference equation, (\ref{Brownian Posterior}), is also the solution to the standard diffusion equation when viewed as transforming \emph{{backwards}} relative to the update time, $T$. As~we shall now discuss, we can promote (\ref{Brownian Posterior}) to a solution to a \emph{{drift}}-diffusion equation by an analogous argument to the one that appeared in Section~\ref{Morris Scheme}. In~particular, we argue that drift in the context of Bayesian diffusion is also associated with a redundancy of description, in~this case related to the specific sequence with which data are~observed.

The reasoning behind this argument is most easily understood through an example. Suppose again that one is performing Bayesian inference on a system  is tthataken to follow a normal distribution with known variance, $\sigma^2$, but~an unknown mean. To~deduce the mean of the distribution governing the system, we observe a sequence of $N$-independent, identically distributed random draws from the true underlying distribution, $E = \{Y_1, ..., Y_N\}$. Starting with a normal prior, one finds that the mean of the posterior distribution after observing the first $n$ pieces of data shall be given by:
\begin{equation}
    \mu(E_n) = \frac{1}{n} \sum_{i = 1}^n Y_i  \, .
\end{equation}

Let $\pi \in \text{Perm}(N)$ be a permutation, and~let $\pi(E) = \{Y_{\pi(1)}, ..., Y_{\pi(N)}\}$ denote the same set of evidence but~in a new order defined by $\pi$. We interpret this transformation as changing the \emph{{sequence}} in which the data are obtained. Crucially, $\pi(E)$ is still a set of $N$-independent, identically distributed random variables drawn from the same data-generating distribution. If~we compute the maximum likelihood estimate based on the first $n$ piece of data appearing in $\pi(E)$, we find:
\begin{equation}
    \pi(\mu(E_n)) = \mu(\pi(E)_n) = \frac{1}{n} \sum_{i = 1}^n Y_{\pi(i)}  \, .
\end{equation}
and hence it is true that $\pi(\mu(E_n)) \neq \mu(E_n)$. However, if~we compute the maximum likelihood estimate on account of all of the data contained in either set, we find:
\begin{equation}
    \pi(\mu(E)) = \frac{1}{N} \sum_{i = 1}^N Y_{\pi(i)} = \frac{1}{N} \sum_{i = 1}^N Y_i = \mu(E)  \, .
\end{equation}
That is, the~terminal maximum likelihood estimate is \emph{{invariant}} with respect to the \emph{{sequence}} in which the data are incorporated into the~model. 

We extrapolate this observation to the statement that the posterior distribution can be assigned an arbitrary path through the space of probability models, provided the terminal distribution remains consistent with the large observation limit, that is, the central tendency towards the data-generating distribution. This is completely analogous to the role of drift in defining an ERG scheme: the invariant definition of an ERG flow is given in terms of the IR fixed point it describes, and the~path through the space of theories by which the theory moves from the UV to the IR is scheme-dependent.

Operationally, we make use of the sequencing freedom in the Bayesian inference to ``seed'' the flow with information about the individual inference trajectory by specifying the $\tau$ path of the maximum a posteriori estimate (MAP). In~the general case where we are given a set of signal parameters $u \in S_U$, we specify the trajectory of the MAP as a flow on the manifold $S_U$ generated by a vector field $V: S_U \rightarrow TS_U$. That is,
\begin{equation}
    \gamma: \mathbb{R} \rightarrow S_U
\end{equation}
such that
\begin{equation}
    \gamma_* \frac{d}{d\tau} = V(\gamma_\tau)
\end{equation}
or, in~more standard notation,
\begin{equation}
    \frac{d}{d\tau} \gamma_{\tau} = \dot{\gamma}_{\tau} = V(\gamma_\tau)
\end{equation}

The trajectory of the MAP arises from maximizing the log-likelihood of the data-generating model. Thus, in~many cases the MAP path, $\gamma_{\tau}$ will realize a gradient descent:
\begin{equation}
    \dot{\gamma}_{\tau} = -\text{grad}_{u} \ln(p_Y(y - G(\gamma_\tau))) = \text{grad}_u \Phi(u;y)\bigg\rvert_{\gamma_{\tau}} = -\mathcal{I}^{ij}(\gamma_{\tau}) \frac{\partial G^k}{\partial u^i} \frac{\partial \Phi}{\partial y^k}\bigg\rvert_{\gamma_{\tau}} \frac{\partial}{\partial u^j} \, ,
\end{equation}
where $\mathcal{I}^{ij}$ are the matrix components of the inverse Fisher metric, and~we have implemented the chain rule to evaluate the derivative. It is very crucial to note that the gradient descent here ensues in the direction of \emph{{increased}} data observation, that is, as~$T \rightarrow \infty$, $\gamma$ approaches the data-generating parameter value, as~desired. 

\subsection{Generic Bayesian Diffusion at Late~T}

The Normal Model is significant because it arises as the late $T$ limit of \emph{{any}} Dynamical Bayesian Inference scheme for which the parameters of the data-generating distribution are equal to some fixed values. This observation arises naturally as an asymptotic limit of the solution to (\ref{Dynamical Bayes Equation}) and~is a statement of the Central Limit Theorem. At~late $T$, the~KL-Divergence can be approximated by the quadratic form:
\begin{equation}
    D_{KL}(u_* \parallel u) = \frac{1}{2} \mathcal{I}_{ij} (u - u_*)^i (u - u_*)^j + \mathcal{O}((\frac{1}{T})^2)
\end{equation}
meaning we can write the unnormalized posterior distribution as:
\begin{equation}
    \varpi_T(u) = e^{-\frac{T}{2} \mathcal{I}_{ij} (u - u_*)^i (u - u_*)^j} \, .
\end{equation}
Provided $\mathcal{I}$ does not depend on $u$, the~normalization is obtained by performing a Gaussian integral, and we can write the posterior distribution as:
\begin{equation} \label{Late T posterior 1}
    \pi_{\tau}(u) = \frac{1}{\sqrt{|2\pi \tau \mathcal{I}^{-1}|}} e^{-\frac{1}{2\tau} \mathcal{I}_{ij} (u - u_*)^i(u- u_*)^j}  \, .
\end{equation}
Here, we have changed the variables to $\tau$ in order to observe that this is the distribution of a multi-variate Brownian~motion. 

Using the gauge freedom discussed in the last section, we can promote this solution to one of the forms:
\begin{equation} \label{Late T posterior 2}
    \pi_{\tau}(u) = \frac{1}{\sqrt{|2\pi \tau \mathcal{I}^{-1}|}} e^{-\frac{1}{2\tau} \mathcal{I}_{ij} (u - \gamma_\tau)^i (u - \gamma_\tau)^j}
\end{equation}
where $\gamma_{\tau}$ is the trajectory of the MAP. As~advertised, both (\ref{Late T posterior 1}) and (\ref{Late T posterior 2}) are consistent with the late $T$ statistics (\ref{Late T behavior}). 

\subsection{A Partial Differential Equation for Bayesian~Inference} \label{Bayesian Diffusion Derivation}

In light of the previous sections, we shall now show that one can derive a convection--diffusion equation describing the evolution of the posterior predictive distribution when it is updated according to Bayes' law. As~we have come to recognize, the~sense in which Bayesian inference describes a diffusion process is in moving \emph{{backwards}} relative to the observation of new information. We will therefore work in the time parameter $\tau$, defined as the inverse to the time parameter $T$ that tracks the amount of data observed. Given the time $\tau$ posterior distribution, which we have argued is of the form of a modified heat kernel, we can define the time $\tau$ posterior predictive model for future data $p_\tau$ by marginalizing over the likelihood model:
\begin{equation} \label{Bayes Kernel}
    p_{\tau}(y) = \int_{S_U} \text{Vol}_U(u) \pi_{\tau}(u) p_{Y\mid U}(y \mid u)    \, . 
\end{equation}

In Section~\ref{SDEs and PDEs}, we reviewed the relationship between the solution to a convention-diffusion equation and the convolution of given boundary value data with a heat kernel. Comparing (\ref{Bayes Kernel}) with (\ref{Markov Operator}), it is natural to regard the posterior distribution as a Markovian transition kernel measuring the probability of going from $y \in S_Y$ to $y - G(u) \in S_Y$. From~this perspective, we define the set of operators $\{P_\tau\}_{\tau \geq 0}$ such that:
\begin{equation}
    P_{\tau}(p_0)(y) := \mathbb{E}_{\Pi_{\tau}}(p_0(y - G(U)))  \, .
\end{equation}

Following the approach outlined in (\ref{Markov Process to PDE}), we can now deduce the diffusion equation generated by the semi-group $\{P_{\tau}\}_{\tau \geq 0}$. By~Taylor Expansion, we can compute:
\begin{equation} \label{Taylor Expansion}
    P_{\tau}(p_0)(y) = \mathbb{E}_{\Pi_\tau}\left(p_0(y) - \frac{\partial p_0}{\partial y^i} \frac{\partial G^i}{\partial u^j} \dot{u}^j \tau + \frac{1}{2} \frac{\partial^2 p_0}{\partial y^i \partial y^j} \frac{\partial G^i}{\partial u^k} \frac{\partial G^j}{\partial u^l} \dot{u}^k \dot{u}^l \tau^2 + \mathcal{O}((\dot{u} \tau)^3)\right) \, .
\end{equation}
Here, we regard $\dot{u}^i \tau$ as corresponding to the infinitesimal flow of the parameter $u^i$ in terms of the vector field $\dot{\gamma}_{\tau}$ generating the flow of the MAP. That is, $\dot{u}^i \tau = \delta u^i$. Following this interpretation:
\begin{equation} \label{P applied to function}
    P_{\tau}(p_0)(y) = \mathbb{E}_{\Pi_{\tau}}\left(p_0(y) - \frac{\partial p_0}{\partial y^i} \frac{\partial G^i}{\partial u^j} \dot{\gamma}_{\tau} \tau + \frac{1}{2} \frac{\partial^2 p_0}{\partial y^i \partial y^j} \frac{\partial G^i}{\partial u^k} \frac{\partial G^j}{\partial u^l} \delta u^k \delta u^l + \mathcal{O}((\delta u)^3)\right) \, .
\end{equation} 

We can now use (\ref{Late T behavior}) to compute the expectation values explicitly to the order specified in the expansion. Because~we are describing the diffusion in terms of the variable $\tau = 1/T$ and we shall eventually be taking the limit as $\tau \rightarrow 0$, we can use the $T \rightarrow \infty$ results described in Equations~(\ref{Late T behavior}) and (\ref{Late T posterior 2}). In~particular, we make use of the fact that
\begin{equation}
    \mathbb{E}_{\Pi_\tau}\left(\prod_{j = 1}^{2n} \delta u^{i_j}\right) = \frac{1}{T^n} \sum_{p \in \mathcal{P}^2_{2n}} \prod_{(r,s) \in p} \mathcal{I}^{i_r i_s} \, .
\end{equation}
Meaning, we can write:
\begin{equation} \label{Quadratic Variation}
    P_{\tau}(p_0)(y) = p_0(y) - \frac{\partial G^i}{\partial u^j} \dot{\gamma}^j_\tau \frac{\partial p_0}{\partial y^i} \tau + \frac{\partial G^i}{\partial u^k} \frac{\partial G^j}{\partial u^l} \mathcal{I}^{kl}(\gamma_\tau) \frac{1}{2} \frac{\partial^2 p_0}{\partial y^i \partial y^j} \tau + \mathcal{O}(\tau^2)  \, .
\end{equation}
We can rewrite these terms in a slightly more illuminating way by writing:
\begin{equation} \label{K matrix}
    \frac{\partial G^i}{\partial u^k} \frac{\partial G^j}{\partial u^l} 
    \mathcal{I}^{kl}(\gamma_\tau) = (G_*\mathcal{I}^{-1})^{ij}\bigg\rvert_{\gamma_\tau} = K^{ij}(\gamma_\tau)  \, .
\end{equation}
Moreover, if~we assume the MAP follows a gradient flow with respect to the log-likelihood, we can also write:
\begin{equation} \label{Y Gradient}
    \frac{\partial G^i}{\partial u^j} \dot{\gamma}^j_\tau = G_{*} \text{grad}_u \Phi(u;y)\bigg\rvert_{u = \gamma_\tau} = -\frac{\partial G^i}{\partial u^j} \frac{\partial G^k}{\partial u^l} \mathcal{I}^{jl} \frac{\partial \Phi}{\partial y^k} = -K^{ij}(\gamma_\tau) \frac{\partial \Phi}{\partial y^j} := m^i  \, .
\end{equation}
Putting everything together, we have therefore shown that:
\begin{equation}
    \frac{P_\tau(p_0)(y) - p_0(y)}{\tau} = -m^i \frac{\partial p_0}{\partial y^i} + K^{ij}(\gamma_\tau) \frac{\partial^2 p_0}{\partial y^i \partial y^j} + \mathcal{O}(\tau^2) 
\, .
\end{equation}
Thus, taking the limit $\tau \rightarrow 0$ we obtain the result:
\begin{equation} \label{Bayesian Diffusion}
    \frac{\partial p_0}{\partial \tau} = \lim_{\tau \rightarrow 0} \frac{P_\tau(p_0) - p_0}{\tau} = -m^i \frac{\partial p_0}{\partial y^i} + K^{ij}(\gamma_\tau) \frac{\partial^2 p_0}{\partial y^i \partial y^j} \, .
\end{equation}
This is precisely of the form of the Kolmogorov equation for a diffusion process with potential $V = \Phi(\gamma_{\tau};y) = \Phi_\tau$! 

It is tempting to interpret the matrix $K^{ij}$ as being associated with the induced Fisher metric in the space of probability distributions along the path of the MAP. From~this perspective, we can regard $Y_{\tau}$, the~random variable associated with the $T$-dependent posterior predictive distribution, as~a stochastic process on a curved space specified by the Stochastic Differential Equation:
\begin{equation} \label{Bayesian SDE}
    dY_\tau = -\text{grad}\;\Phi_\tau \; d\tau + dW_\tau \, .
\end{equation}
The potential is given by $\Phi_\tau$, which is minus the log-likelihood associated with the time $\tau$ maximum likelihood estimate for the generating distribution. Thus, we have shown that a Dynamical Bayesian inference induces a gradient flow with respect to the log-likelihood of the data-generating~distribution.   

\section{The ERG flow/Dynamical Bayesian Inference~Correspondence} \label{ERG = Backwards Inference}

We are now prepared to build the dictionary relating ERG flow and Bayesian Inference. For~simplicity, we shall consider one-parameter families of probability distributions on finite dimensional sample spaces; however, it is a simple exercise to generalize these insights to the infinite dimensional case as well. To~begin, let us review the work we have presented in the previous~sections. 

In Section~\ref{ERG is Diffusion}, we recalled the Wegner--Morris formulation for ERG and demonstrated that it is equivalent to a one-parameter family of probability distributions described by a gradient flow with respect to the relative entropy:
\begin{equation} \label{Wegner Morris for ERG}
    \frac{\partial p_t}{\partial t} = -\text{grad}_{\mathcal{W}_2} D_{KL}(p_t \parallel q_t) = \frac{\partial}{\partial x^i}\left(p_t g^{ij} \frac{\partial}{\partial x^j} \Sigma\right) = \frac{\partial}{\partial x^i}\left(p_t \Psi^i\right) \, .
\end{equation}

In this equation, we interpret the time parameter $t$ as a logarithm of the scale -- $t = \ln\Lambda$. The~distribution $q_t = \hat{q}_t/Z_q$, with~$\hat{q}_t = e^{-V}$ for the potential $V$, can be regarded as specifying the ERG scheme through the fixing of a stationary point along the flow generated by (\ref{Wegner Morris for ERG}). The~choice of $V$ \emph{{defines}} the functional, $\Sigma = -\ln(\frac{\hat{p}_t}{\hat{q}_t})$, and~the reparameterization kernel, $\Psi = \text{grad}_g \Sigma$; hence, it is equivalent to a choice of scheme in the standard Wegner--Morris sense. As~we have reviewed in Section~\ref{SDEs and PDEs}, the~Fokker--Planck equation is equivalent to the Kolmogorov Forward equation for the Stochastic Process governed by the Stochastic Differential Equation:
\begin{equation}
    dX_t = -(\text{grad}_g V) dt + \sqrt{2} dW_t  \, .
\end{equation}
This is intuitively satisfying since such an equation describes a stochastic gradient descent of the potential function $V$. Once initial data are supplied in the form of a UV theory, $p_0$, (\ref{Wegner Morris for ERG}) completely describes an ERG flow terminating at an IR fixed~point. 

In Section~\ref{Dynamical Bayes}, we introduced the notion of Dynamical Bayesian Inference. Dynamical Bayesian Inference describes a one-parameter family of probability distributions obtained by implementing Bayes' law using data collected from a continuous time stochastic process. To~quantify this family of distributions, we introduced the ``time'' parameter, $T$, which corresponds to the number of data incorporated into the model. In~the direction of increasing $T$, the~inferred probability model converges onto the distribution generating the observed data. In~this respect, a~Dynamical Bayesian flow has the complexion of an \emph{{inversion}} in the ERG sense because it begins with an uninformed prior distribution but eventually converges to an informed distribution. In~the language of ERG, this describes a flow from an IR theory to a UV theory. The~pair consisting of an uniformed prior, along with the specification of a sufficiently complete set of data, therefore defines a flow terminating with a UV theory. We take this as our definition of a Dynamical Bayesian~Flow. 

This picture suggests that Dynamical Bayesian Inference and ERG flow are {\it{{inverse}}} of each other. If~we can find a Dynamical Bayesian inference that begins with a prior distribution equal to the IR fixed point of an ERG flow, and~which terminates at a data-generating distribution equal to the UV initial data of said ERG flow, we will have obtained an inversion of the ERG flow. Let us now describe a strategy for determining pairs of ERG flows with Dynamical Bayesian~Flows. 

\subsection{Dynamical Bayesian Flow $\rightarrow$ ERG~Flow}

Suppose we are given a Dynamical Bayesian flow and asked to determine an ERG flow that is inverse to it. Recall, we have defined a Dynamical Bayesian Flow as the pair, $(q_0, \{Y_t\}_{t = 1}^T)$, where $q_0$ is an uninformed prior distribution and~$\{Y_t\}_{t = 1}^T$ is a set of data generated by a distribution $p_*$. Using these data, it is straightforward to \emph{{define}} the corresponding ERG flow as the one-parameter family of probability distributions obtained from the Dynamical Bayesian flow when viewed as evolving \emph{{backwards}} with respect to the time parameter $T$. In~other words, one takes the terminal distribution of the Dynamical Bayesian flow, $p_*$, as~defining the UV initial data of the ERG flow and~obtains subsequent probability distributions along the ERG flow by removing items of data from the inferred~model.

In Section~\ref{Bayesian Diffusion Derivation}, we derived a partial differential equation governing just such a situation, in~which the posterior predictive distribution evolves \emph{{backwards}} against the collection of new data. The~resulting partial differential equation describes a diffusion process that we dubbed \emph{{Bayesian Diffusion}}: that is, one obtains a one-parameter family of probability distributions $\{p_{\tau}\}$ such that $p_0 = p_*$ and for which:
\begin{equation}
    \frac{\partial p_{\tau}}{\partial \tau} + m^i \frac{\partial p_{\tau}}{\partial y^i} - K^{ij}(\gamma_{\tau}) \frac{\partial^2 p_{\tau}}{\partial y^i \partial y^j} = 0 \, .
\end{equation}
This differential equation describes the evolution of a probability model governing the stochastic process $Y_{\tau}$, which itself is governed by the Stochastic Differential Equation
\begin{equation}
    dY_{\tau} = -(\text{grad}_{K} \Phi) d\tau + dW_{\tau} \, .
\end{equation}
Here, $\gamma_{\tau}$ is a trajectory in the model space of the theory describing the path of the maximum a posteriori parameter estimate generated by the sequence of observed data, $K$ is the pullback of the Fisher Information Metric on the space of models defined in (\ref{K matrix}), and~$\Phi$ is the log-likelihood function associated with the Bayesian Inference~scheme. 

Together, the~aforementioned items constitute a \emph{{choice of scheme}} for the Dynamical Bayesian flow. Bayesian diffusion can therefore be associated with an ERG flow governed by the Wegner--Morris equation:
\begin{equation} \label{Bayesian WM Flow}
    \frac{\partial p_{\tau}}{\partial \tau} = -\text{grad}_{\mathcal{W}_2} D_{KL}(p_{\tau} \parallel q_{\tau}) = \frac{\partial}{\partial y^i}\left(p_{\tau} K^{ij} \frac{\partial}{\partial y^j} \Sigma\right) = \frac{\partial}{\partial y^i}\left( p_{\tau} \Psi^i\right)
\end{equation}
where now the stationary distribution, scheme function, and~reparameterization kernel are, respectively, given by:
\begin{equation}
    q_{\tau} = \frac{e^{-\Phi(\gamma_{\tau};y)}}{Z_{q}}; \;\;\; \Sigma = \Phi(\gamma_{\tau};y) - \Phi(u;y); \;\;\; \Psi = \text{grad}_{K} \Sigma \, .
\end{equation}
These data, along with the distribution $p_*$, define an ERG flow in the Wegner--Morris sense, which, by~construction, is the inverse of the Dynamical Bayesian flow we began~with. 

\subsection{Dynamical Bayesian Flow $\leftarrow$ ERG~Flow}

Conversely, suppose that we are given an ERG flow in a Wegner--Morris form and asked to determine a Dynamical Bayesian flow that is inverse to it. To~do so, we first identify the Bayesian Diffusion process that the ERG flow corresponds to. Since we have shown that ERG and Bayesian Diffusion are governed by the same equations, this is as simple as translating between the ERG scheme and the Dynamical Bayesian scheme. Comparing (\ref{Wegner Morris for ERG}) and (\ref{Bayesian WM Flow}), we deduce that the ERG associated with an optimal transport with potential $V$ and sample space metric $g$ is equivalent to a Bayesian diffusion in which $V$ is taken as the log-likelihood function, and~$g$ is identified with the metric $K$. These data are sufficient to define a Bayesian inference problem. Notice, if~we go all the way back to (\ref{Seed Action}), we can finally provide a conceptual understanding of the seed action: the seed action sets the log-likelihood for the Bayesian Inference scheme related to the ERG flow it~defines. 

\subsection{A~Dictionary}

We summarize the analysis of this final section in Table~\ref{OT ERG DB}. It provides a dictionary translating between the Wegner--Morris equations relevant to finite sample space ERG, infinite dimensional sample space ERG, and~Bayesian~Diffusion. 

\begin{table}[H]
\centering
\begin{tabular}{llll}
                                           & $\textbf{Finite Dim. ERG}$ & $\textbf{Infinite Dim. ERG}$                                                                                                                                                                                                                                       & $\textbf{Bayesian Diffusion}$ \\
$\textbf{Time Parameter}$                  & $t$                          & $\ell = \ln(\Lambda)$                                                                                                                                                                                                                                & $\tau = \frac{1}{T}$                         \\
$\textbf{Metric}$                          & $g^{ij}$                     & $\dot{C}_{\Lambda}(x,y)$                                                                                                                                                                                                                             & $\mathcal{I}^{ij}(\gamma_{\tau})$            \\
$\textbf{Potential}$                       & $V$                          & $-2\hat{S}_{\Lambda}[\phi]$                                                                                                                                                                                                                          & $\Phi(\gamma_{\tau};y)$                      \\
$\textbf{Scheme}$                          & $\Sigma = -\ln(\frac{p}{q})$ & $\Sigma = S_{\Lambda}[\phi] - 2\hat{S}_{\Lambda}[\phi]$                                                                                                                                                                                              & $\Sigma = \Phi(u;y) - \Phi(\gamma_{\tau};y)$ 
\end{tabular}
\caption{Dictionary relating ERG and Bayesian Diffusion.}
\label{OT ERG DB}
\end{table}

\subsection{Renormalizability and~Scale}

The dictionary in Table~\ref{OT ERG DB} provides a blueprint for interpreting ERG in the language of Statistical Inference. In~this role, our work suggests new approaches to understanding and resolving many interesting problems inside and outside of field theory. As~a demonstration, we shall use this final section to discuss the meaning of renormalizability in the context of an ERG flow related to the Bayesian Inference~paradigm. 

An important observation in (\ref{OT ERG DB}) is the correspondence between the Fisher Metric in the inference context and $\dot{C}_{\Lambda}(x,y)$ in the ERG context. In~the exact renormalization of a free field theory, $\dot{C}_{\Lambda}(x,y)$ is the regulated two-point function and~therefore sets a running momentum scale for operators in the theory. The~Fisher Metric, $\mathcal{I}$, plays an analogous role in the Bayesian Inference scheme as a generalized two-point function encoding a notion of \emph{{scale}} through the covariance between~operators. 

The interpretation of the Fisher Metric as defining an energy scale is made very clear when we consider the inverse Bayesian flow as a diffusion process. Let $\mathcal{M} = \text{dens}(S)$ denote the manifold of probability distributions over a sample space $S$. Then, a~Bayesian diffusion, or~equivalently an Exact Renormalization Group flow, can be described by a drift-diffusion process generated by an operator $L: \mathcal{M} \rightarrow \mathcal{M}$, such that:
\begin{equation} \label{Diffusion}
    \left(\frac{\partial}{\partial t} + L\right) p_t = 0 \, .
\end{equation}
Given initial data ({for example, in~an ERG we would provide a $UV$ theory}) $p_0 \in \mathcal{M}$, we can write the solution to (\ref{Diffusion}) symbolically as:
\begin{equation} \label{Solution to Heat Equation}
    p_t = e^{-tL}\left(p_0\right) \, .
\end{equation}
where $e^{-tL}$ is the heat kernel of $L$. To~give more concrete meaning to (\ref{Solution to Heat Equation}), let us assume that $L$ is a positive definite operator that can be diagonalized as
\begin{equation}
    L(\psi_n) = \lambda_n^2 \psi_n \, .
\end{equation}
Here, $\{\psi_n\}$ is a countably infinite set forming a basis for $\mathcal{M}$, and~$\lambda_n$ is non-negative but~may be equal to zero. An~arbitrary element $p \in \mathcal{M}$ can be expanded as a series: $p = \sum_n p^n \psi_n$, where $p^n$ is the coordinate of $p$ in the eigendirection $\psi_n$. 

The spectrum of $L$ defines an \emph{{emergent energy scale}} in the following sense: consider the action of (\ref{Solution to Heat Equation}) as given by:
\begin{equation} \label{Heat Kernel applied to Basis}
    p_t = \sum_n e^{-t\lambda_n^2} p_0^n \psi_n  \, .
\end{equation}
(\ref{Heat Kernel applied to Basis}) dictates that the projection of $p_t$ onto each mode, $\psi_n$, is damped over time with a strength determined by on the ``energy'' $\lambda_n^2$. At~time $t$, the~effective description exponentially suppresses modes that have large eigenvalues relative to the operator $L$ and~thus sequentially removes these modes in a generalized integration over effective ``momentum shells''. From~Equation~(\ref{Bayesian Diffusion}), we can see that the operator $L$ is generically a convection--diffusion operator with a diffusion matrix given by the Fisher Information. Through further comparison to (\ref{Polchinski Heat Equation}), we see that in ERG for physical theories the analogous role is played by the regulated two-point function. This leads to the important conclusion that, in~physical contexts, the~emergent energy scale is in fact equivalent to a physical~one. 

More generally, interpreting the two-point function, or~the covariance matrix, in~a statistical inference problem as generating an emergent energy scale provides the foundation for an information-theoretic interpretation of the conditions for renormalizability. In~Wilsonian RG, a~theory is said to be non-renormalizable if the divergences present in higher-order Feynman diagrams can only be canceled by the introduction of an infinite number of arbitrarily high-energy couplings~\cite{parisi1975theory}. By~contrast, a~theory is said to be renormalizable if arbitrarily higher-order Feynman diagrams can be computed by introducing only a finite number of operator sourced~counterterms. 

In the language of statistical inference, the operator content of a theory is related to the problem of model selection~\cite{anderson1998comparison,balasubramanian1997statistical}. In~the context of parametric statistics ({as we have mentioned in the introduction, the~relationship between Wilsonian RG and ERG is analogous to the relationship between Parametric and Non-Parametric statistics}), model selection can be reduced to determining the set of sufficient parameters needed to form a model that can accurately compute the expectation values of any observable associated with the system of interest. A~natural framing of this problem is given in terms of $n$-point correlation functions for the random variable, $Y$, observed throughout the inference. It is sensible to restrict our attention to $n$-point functions since arbitrary observables can be constructed from them using a Taylor expansion. Put differently, a~probability distribution can be reconstructed with knowledge of all its~moments. 

In view of the previous discussion, higher $n$-point functions can be interpreted as encoding information at higher values of the emergent energy scale. This inspires the interpretation that an inference model is ``renormalizable'' if a finite $N$ exists such that for any $n > N$, the~$n$-point function can be computed with the information contained only in $m$-point functions with $m \leq N$. In~other words, an energy as measured through $L$, $\lambda_*^2$ exists, above~which all of the information in the theory is actually encoded in lower energy operators. This happens, for~example, in~a Gaussian theory in which all $n$-point functions higher than $N = 2$ can be formulated as sums of products of $2$-point functions using Wick's theorem. If~no such finite $N$ exists, the~inference problem is ``nonrenormalizable''. A~nonrenormalizable theory can therefore be understood as a theory in which an infinite number of $n$-point functions will be required to compute the expectation values of arbitrary observables. In~other words, there is no energy scale above which information becomes encoded in the energy scales below it---every energy scale contributes, in~some sense, independently to the~theory.

\section{Discussion} \label{Discussion}

In this note, we have demonstrated that an ERG flow can be identified with a diffusion process that is \emph{{inversely}} related to a Dynamical Bayesian Inference scheme. In~particular, we have argued that ERG flow can be understood as a one-parameter family of probability distributions arising where data are continuously removed from the inferred probability model. We have motivated this interpretation by illustrating that the equations governing ERG and Bayesian Diffusion can be brought into direct correspondence with one another, as~outlined in Section~\ref{ERG = Backwards Inference}. The~resulting dictionary provides a novel, fully information-theoretic language for understanding ERG flow. It also provides an operational answer to the question of what it means to ``invert'' an ERG~flow.

From a very general perspective, the~solution to this problem can be framed in the following way. Given a preliminary probability distribution, $p_0$, we imagine running our model through a noisy channel generated by a diffusion operator $B$. In~other words, we produce a probability distribution, $p_{\tau}$, which solves the differential equation:
\begin{equation} \label{Diffusion Phase}
    \frac{\partial p_{\tau}}{\partial \tau} + B (p_{\tau}) = 0 \, ,
\end{equation}
with initial data $p_0$. After~a given period of time, $t$, we obtain a new probability distribution, $p_{t} = e^{-t B} (p_0)$, which has lost some of the information previously contained in $p_0$ to diffusion. In~the case of an ERG flow viewed from the functional diffusion perspective, we can regard this loss of information as being generated by a coarse-graining scheme encoded in the operator $B$. 

We then ask the question, can this diffusion process can be ``inverted''? Since exact inversion may not be possible, we frame this problem in the form of an optimization scheme. Consider the set $\mathcal{F}$ consisting of all operators $F$, generating a one-parameter flow of probability distributions, $q_{T}$, such that:
\begin{equation}
    \frac{\partial q_{T}}{\partial T} + F (q_{T}) = 0 \, .
\end{equation}
subject to an initial condition $q_0$. If~we take the initial data of this process to be the terminal distribution of the diffusion process given by (\ref{Diffusion Phase}), $q_0 = e^{-tB}(p_0)$, we can interpret the solution $q_t = e^{-tF}(q_0)$ as a \emph{{reconstruction algorithm}} for the initial data, $p_0$. It is natural to identify the optimal reconstruction algorithm as the operator $F_* \in \mathcal{F}$, for which the relative entropy between the reconstructed distribution, $q_t$, and~the initial data, $p_0$, is minimal:
\begin{equation} \label{Reconstruction Problem}
    F^* := \argmin_{F \in \mathcal{F}} D_{KL}(p_0 \parallel e^{-tF} \circ e^{-tB} p_0) \, .
\end{equation}

In this language, we interpret the main result of our paper as dictating that, given an ERG generated by a diffusion operator $B$, the~optimal reconstruction operator $F^*$ corresponds to a continuous Bayesian Inference scheme in which the information lost to coarse-graining is \emph{{re-learned}} and~hence reincorporated into the model. This clarifies the sense in which an ERG is ``invertible'' as long as we allow for the \emph{{reconstruction}} of information ostensibly destroyed by~diffusion. 

One can visualize this process as follows: imagine an experimenter performing a statistical inference experiment in  which they observe a collection of data $\{Y_i\}_{i = 1}^T$, generated from the distribution $p_0$. Next, imagine we can place each of the observations along the real axis, distinguishing a series of points, each of which we label by a probability distribution $\{p_T\}_{T = 0}^{\infty}$. The~probability distribution at the $T$th point, $p_T$, is obtained by incorporating all of the data to the left of $T$ into a model using Bayes' law. Moving to the \emph{{left}} along this axis corresponds to \emph{{dis}}incorporating data from the model and~therefore induces a diffusion process and by extension an ERG scheme. Conversely, moving to the \emph{{right}} along this axis corresponds to \emph{{re}}incorporating lost data and~therefore \emph{{inverts}} the ERG~flow.

Framing the relationship between ERG and Statistical Inference in terms of the reconstruction problem (\ref{Reconstruction Problem}) suggests several interesting paths for future study. Firstly, the~reconstruction problem is equivalent to a common problem encountered in Machine Learning when one wishes to sample data from an analytically intractable distribution, $p_0$. An~approach to this problem goes by the name of \emph{{Diffusion Learning}} \cite{sohl2015deep,neal2001annealed, ramesh2021zero,ramesh2022hierarchical}. Diffusion learning is a two-step process: first, one uses a diffusion operator, $B$, with~a known fixed point to transform the initial data $p_0$ into an analytically tractable form. Then one identifies a second diffusion operator, $F$, which optimally reconstructs the initial data without sacrificing analytic tractability. This routine is equivalent to (\ref{Reconstruction Problem}), provided we restrict the set of allowed reconstruction algorithms to operators that generate diffusion processes. More generally, the~information-theoretic formulation of ERG constructed in this paper renders renormalization in a form that is amenable to applications outside of pure physics. We hope this will catalyze continued work, especially at the intersection of physics and data science, geared towards constructing and better understanding machine learning algorithms like diffusion learning. {Since the original drafting of this note, some work to this end has been undertaken. In~\cite{Berman:2023rqb}, the approach introduced in this paper was adapted into a practical renormalization scheme for generic statistical inference models including neural networks. As~a proof of concept, this scheme was subsequently applied to construct renormalization group flows for autoencoder neural networks}.

A second fascinating implementation of (\ref{Reconstruction Problem}) appears in the study of Holography~\cite{maldacena1999large}. There, one is interested in \emph{{reconstructing}} a bulk spacetime from the data contained in a quantum field theory on its conformal boundary~\cite{dong2016reconstruction,pastawski2015holographic}. The~relationship between our work and bulk reconstruction is very natural. In~modern literature, bulk reconstruction is often interpreted through the language of Quantum Error Correction as the inversion of a quantum channel associated with the propagation of bulk data into a subregion of the conformal boundary~\cite{lashkari2016canonical,cotler2019entanglement, faulkner2020holographic,furuya2020real}. Placed in this context, the~bulk reconstuction problem is a non-commutative generalization of (\ref{Reconstruction Problem}) in which one replaces probability distributions by density operators, and~the maps $B$ and $F$ by Quantum Channels~\cite{ohya2004quantum,junge2018universal} ({see Appendix~\ref{Petz and Bayes} for a short discussion of the relationship between this paper and error correction}). This suggests that by studying a quantum version of the correspondence introduced in this paper, one might be able to shed light on some of the mysterious aspects of the $AdS/CFT$ correspondence. Beyond~this, there are also interesting questions that pertain to the generalization of our correspondence into the general language of non-commutative probability theory, $C^*$ operator algebras. These include understanding statistical inference for non-commutative operator algebras~\cite{helstrom1969quantum,beny2015information,beny2015renormalization}, the~study of non-commutative diffusion processes~\cite{carlen2014analog,carlen2017gradient,carlen2020non}, and~the formulation of renormalization techniques that work beyond the scope of Euclidean QFT in the vein of entanglement renormalization and tensor networks~\cite{nozaki2012holographic,swingle2012entanglement,alvarez1999geometric,leigh2014holographic,mollabashi2014holographic,evenbly2015tensor}.

Finally, our picture of ERG can provide a powerful tool for constructing and interpreting theorems about renormalization. By~formulating ERG flows as a \emph{{related}} statistical inference problem, one obtains a stark accounting of the information/degrees of freedom contained in the renormalized theory at any point over the course of its flow. Such knowledge is of great use in computing the information lost between various points along an RG flow and, especially, in~constructing and interpreting RG monotones~\cite{zamolodchikov1986irreversibility,myers2010seeing, casini2007c, casini2015mutual,casini2017markov}. We are hopeful that our conceptual approach to the ERG can expand renormalization in its role as a toolset for studying the space of Quantum Field~Theories.

\section{Acknowledgements}

We thank Jonathan Heckman for collaboration on dynamical Bayes in \cite{berman2022dynamics} which led to many of the ideas in this paper. We are grateful to the participants of the ``String Data 2022" conference where this work was first presented and subsequent comments from Semon Rezchikov and Miranda Cheng. We also wish to thank Alex Stapleton for related collaboration on forth-coming work on diffusion models and reconstruction channels in holography. Finally, we thank Samuel Goldman and Robert Leigh for enlightening discussions on Exact Renormalization. 
DSB acknowledges support from Pierre Andurand over the course of this research. MSK is supported through the Physics department at the University of Illinois at Urbana-Champaign.

\pagebreak

\begin{appendix}

\section{Optimal Transport} \label{Appendix Optimal Transport}

Optimal transport consists in redistributing the mass between two probability measures in order to accomplish a cost minimization. To be precise, let $Y \in S$ be a random variable in the sample space $S$. We regard $S$ as an orientable differentiable manifold possessing a reference measure, $\text{Vol}_S(y)$, which is simply the volume form on $S$. A probability measure can then be obtained by considering a measurable function: $p: S \rightarrow \mathbb{R}_+$ which is normalized in the integral sense:

\begin{equation}
    \int_S \star p = \int_S p(y) \text{Vol}_S(y) = 1
\end{equation}

Optimal transport can then be stated in two equivalent forms. The \emph{Monge Formulation} is as follows: Given a space $S$ and two probability distributions $p_1$ and $p_2$ corresponding to measures $\mu_1 = \star p_1$ and $\mu_2 = \star p_2$ we seek a \emph{transport function} $T: S \rightarrow S$ such that:

\begin{enumerate}
    \item \begin{equation}
    T^* \mu_2 = \mu_1
\end{equation}
meaning the transport function pulls back mass from $\mu_2$ to $\mu_1$, or
\begin{equation}
    \int_{T(U)} \mu_2 = \int_U \mu_1
\end{equation}
for any subset $U \subset S$. Note, this is usually written in terms of the ``pushforward": $T_{\#} \mu_1 = \mu_2$ which is the pullback by the inverse map $T^{-1}: S \rightarrow S$ i.e. $\mu_2 = (T^{-1})^* \mu_`$. 
    \item The transport function is selected to minimize the objective function:
    \begin{equation}
    M[T] = \int_S \mu_1(y) c(y, T(y))
\end{equation}
where $c: S \times S$ is some cost function which is typically associated with a distance on the sample space. 
\end{enumerate}

Alternatively, we can state the optimal transport problem in the \emph{Kantorovich Formulation}. In that case one begins with a joint measure $\Pi$ on the sample space $S \times S$ that pushes forward (in the sense of integrating along the fiber, or simply marginalizing in the probabilistic sense) to the measure $\mu_1$ and $\mu_2$ respectively. That is: $\Pi(y_1,y_2) = \pi(y_1, y_2) \text{Vol}_S(y_1) \wedge \text{Vol}_S(y_2)$ 

\begin{enumerate}
    \item \begin{equation}
        \mu_1(y_1) = \text{Vol}_S(y_1) \wedge \int_{S} \text{Vol}_S(y_2) \pi(y_1, y_2) \;\;\;\;\; \mu_2(y_2) = \text{Vol}_S(y_2) \wedge \int_S \text{Vol}_S(y_1) \pi(y_1, y_2)
    \end{equation}
    We shall henceforth denote by $\Gamma(\mu_1, \mu_2)$ the set of joint measures which marginalize to $\mu_1$ and $\mu_2$. 
    \item The joint measure $\Pi$ is chosen so as to minimize the joint expectation value for the cost:
    
    \begin{equation}
        K(\Pi) = \int_{S \times S} \Pi(y_1, y_2) c(y_1, y_2)
    \end{equation}
\end{enumerate}
Notice, if we choose $\pi(y_1, y_2) = p_1(x) \delta(y - T(x))$ the Kantorovich objective function is equivalent to the Monge function. 

There is an important theorem which states that for the $\ell^2$ cost function, $c(y_1,y_2) = \norm{y_1 - y_2}^2$, there exists a smooth solution to the Monge problem. In particular, there will exist a smooth function $f: S \rightarrow \mathbb{R}$ for which

\begin{equation} \label{Monge Solution}
    T^i(y) = g^{ij}(y) \frac{\partial}{\partial y^j} f(y)
\end{equation}
or
\begin{equation}
    T(y) = \text{grad} f(y)
\end{equation}
Recall the relationship between $\mu_1$ and $\mu_2$: $\mu_1 = T^* \mu_2$ or

\begin{equation}
    \mu_1 = T^*(\mu_2) = \det\left(\frac{\partial T^i(y)}{\partial y^j}\right) p_2(T(y)) \text{Vol}_S(y)
\end{equation}
or, with respect to the probability distributions:

\begin{equation}
    p_1(y) = \det\left(\frac{\partial T^i}{\partial y^j}\right) p_2(T(y))
\end{equation}
Hence, with respect to the solution we have specified in (\ref{Monge Solution}) we can write:

\begin{equation}
    p_1(y) = \det\left(\frac{\partial}{\partial y^j} g^{ik}(y) \frac{\partial}{\partial y^k} f \right) p_2(\text{grad} f(y))
\end{equation}
or, put more suggestively:
\begin{equation}
    \frac{p_1(y)}{p_2(\text{grad}f(y))} = \Delta f(y)
\end{equation}
where here $\Delta$ is the Laplacian. 

The \emph{Wasserstein Distance} is a metric on the space of probability measures which is defined as the solution to an optimal transport problem with a specified cost:

\begin{equation}
    \mathcal{W}_c(\mu_1, \mu_2) := \text{inf}_{\Pi \in \Gamma(\mu_1, \mu_2)} \int_{S \times S} \Pi(y_1, y_2) c(y_1, y_2)
\end{equation}
Of particular interest to us will be the Wasserstein two distance, which is the Wasserstein distance defined with respect to the $\ell^2$ cost function:

\begin{equation}
    \mathcal{W}_2(\mu_1, \mu_2) = \left(\text{inf}_{\Pi \in \Gamma(\mu_1, \mu_2)} \int_{S \times S} \Pi(y_1, y_2) \norm{y_1 - y_2}^2\right)^{1/2}
\end{equation}

\section{ERG and Error Correction} \label{Petz and Bayes}

In this section, we would like to note how the work we have presented in this paper connects to a related approach to understanding RG through the language of quantum error correction \cite{cotler2019entanglement,faulkner2020holographic,furuya2020real}. 

Given an operator algebra, $\mathcal{A}$, corresponding to a set of observable degrees of freedom, a theory can be thought of as a state, or, in more physical language, as a density operator, which assigns to each operator an expectation value. We shall denote the set of states on $\mathcal{A}$ as $\mathcal{A}_*$. A \emph{quantum channel}, $\mathcal{E}: \mathcal{A}_* \rightarrow \mathcal{B}_*$, is a completely positive, trace preserving, linear map from states on an operator algebra $\mathcal{A}$ to states on an operator algebra $\mathcal{B}$. A quantum channel is a natural mathematical representation for the generator of an RG flow because of the \emph{Data Processing Inequality}:

\begin{equation} \label{Data Processing Inequality}
    D_{KL}(\rho \parallel \rho') \geq D_{KL}(\mathcal{E}(\rho) \parallel \mathcal{E}(\rho'))
\end{equation}
One can interpret (\ref{Data Processing Inequality}) as stating that the distinguishability of states is decreased under the action of any quantum channel. For this reason, a quantum channel is sometimes also referred to as a \emph{coarse-graining map}, in analogy with an RG flow.   

A quantum channel is \emph{exactly reversible} if and only if it is \emph{sufficient} in the sense that no information is lost: that is for all $\rho, \rho' \in \mathcal{A}_*$ the data processing inequality is saturated and

\begin{equation} \label{Sufficiency Condition}
    D_{KL}(\rho \parallel \rho') = D_{KL}(\mathcal{E}(\rho) \parallel \mathcal{E}(\rho'))
\end{equation}
In this case, there will exist a complimentary channel, $\mathcal{P}_{\rho', \mathcal{E}}: \mathcal{B}_* \rightarrow \mathcal{A}_*$, called the \emph{Petz Map} such that $\mathcal{P}_{\rho', \mathcal{E}} \circ \mathcal{E}(\rho) = \rho$. 

In general, (\ref{Sufficiency Condition}) will only be met for a subset of states, $\mathcal{C}_* \subset \mathcal{A}_*$. The operator algebra associated with this set of states is denote by $\mathcal{C}$, and called the \emph{Code Subspace}.  The operators which live inside the code-subspace are defined by the property that their expectation values are invariant under the flow generated by $\mathcal{E}$. Leveraging the interpretation of a quantum channel as generating an \emph{RG Flow}, we can therefore interpret the code subspace of $\mathcal{E}$ as corresponding to the set of relevant operators. 

In this paper we have provided a concrete link between information theory and ERG through the intermediary of statistical inference by applying the concept of Bayesian inversion to the heat flow describing an ERG. Regarding Bayesian Inference as dual to ERG fits neatly into the Error Correcting picture discussed above. The Petz Map of a quantum channel $\mathcal{E}$ can be defined as its formal adjoint with respect to a non-commutative generalization of the Fisher Information Metric on $\mathcal{A}_*$.\footnote{$g_{\rho}: T_{\rho} \mathcal{A}_* \times T_{\rho} \mathcal{A}_* \rightarrow \mathbb{R}$, such that $g_{\rho}(X, Y) = \text{Tr}(X \Omega_{\rho}^{-1} (Y))$, where here $\Omega_{\rho}^{-1}$ is a map that corresponds to a non-commutative generalization of ``division by $\rho$". Explicitly,

\begin{equation}
    \Omega^{-1}_{\rho}(Y) = \frac{d}{dt}\bigg\rvert_{t = 0}\log(\rho + tY)
\end{equation} 

The metric $g_{\rho}$ takes the schematic form, $\text{Tr}(\frac{XY}{\rho})$, which is equivalent to the usual form of the Fisher Metric.}  
\begin{equation} \label{Non-Commutative Bayes}
    g_{\rho}(X, \mathcal{P}_{\rho, \mathcal{E}}(Y)) = g_{\mathcal{E}(\rho)}(\mathcal{E}(X), Y)
\end{equation}
When the operator algebra in question is \emph{commutative} the set of states becomes equivalent to the space of probability distributions, and the metric $g_{\rho}$ reduces to the unique Fisher Metric on this space. A consequence of this fact is that the Petz Map for commutative operator algebras as obtained through (\ref{Non-Commutative Bayes}), is \emph{equivalent} the Bayesian posterior with prior $\rho$. We prove this statement now:

Let $\mathcal{A}$ be a commutative algebra. For example, its elements may correspond to the set of measurable functions on a domain, $S$, or, in the parlance of Probability Theory, random variables on the sample space $S$. A state on $\mathcal{A}$ is then a probability measure on $S$, which is a measurable function $p: S \rightarrow \mathbb{R}$ that is normalized in the integral sense:

\begin{equation}
    \text{Tr}_{\text{Vol}}(p) = \int_S \text{Vol}_S(x) p(x) = 1
\end{equation}
Here $\text{Vol}_S$ is the reference measure on $S$, if $S$ is an orientable differentiable manifold this is nothing but the volume form. The pairing of a state, $p \in \mathcal{A}_*$ and a random variable $f \in \mathcal{A}$ is the computation of an expectation value:

\begin{equation}
    p[f] := \mathbb{E}_{p}(f(X)) = \text{Tr}_{\text{Vol}}(pf) = \int_{S} \text{Vol}_S(x) p(x) f(x)
\end{equation}

A channel between states on commutative algebras, $\mathcal{E}: \mathcal{A}_* \rightarrow \mathcal{B}_*$, is a \emph{Stochastic Map}. Let $S_A$ and $S_B$ denote the sample spaces associated with the algebras $\mathcal{A}$ and $\mathcal{B}$, respectively. Then, we can associate to $\mathcal{E}$ a conditional probability distribution: $p_{B \mid A}: A \times B \rightarrow \mathbb{R}$ such that:

\begin{equation}
    \int_{S_B} \text{Vol}_B(y) p_{B \mid A}(y \mid x) = 1; \;\;\; \forall_{x \in S_A}
\end{equation}
and
\begin{equation}
    \int_{R \subset S_B} \text{Vol}_{B}(y) p_{B \mid A}(y \mid x) = \mathbb{P}(y \in R \mid X = x)
\end{equation}
More to the point, given a marginal probability distribution $p_A \in \mathcal{A}_*$, the action of the map $\mathcal{E}$ is given by the following integral:

\begin{equation}
    \mathcal{E}(p_A) = \int_{S_A} \text{Vol}_A(x) p_{B \mid A}(y \mid x) p_A(x) \in \mathcal{B}_*
\end{equation}

The Fisher Metric on $\mathcal{A}_*$ takes the form:\footnote{This form of the Fisher Metric should be compared with the more standard form:

\begin{equation}
    (g_{p_A})_{ij} = \int_{S_A} \text{Vol}_A(x) p_A(x \mid \theta) \frac{\partial \log(p_A(x \mid \theta))}{\partial \theta^i} \frac{\partial \log(p_A(x \mid \theta))}{\partial \theta^j} = \int_{S_A} \text{Vol}_{A}(x) \frac{1}{p_A(x \mid \theta)} \frac{\partial p_A(x \mid \theta)}{\partial \theta^i} \frac{\partial p_A(x \mid \theta)}{\partial \theta^j}
\end{equation}
Here $\theta$ are a set of parameters specifying a probability density in a parametric family. In the first equality, one regards $\ell_i = \frac{\partial \log(p_A(x \mid \theta))}{\partial \theta^i}$ as a basis for $T_{p_A}\mathcal{A}_*$, with $T_{p_A}\mathcal{A}_*$ identified as the set of random variables on $S_A$ with zero expectation value. The second equality arises form a simple algebraic manipulation of the first, but implies a different condition on $T_{p_A}\mathcal{A}_*$, namely that is the set of measurable functions on $S_A$ that integrate to zero. The second condition is consistent with our chosen specification of $T_{p_A}\mathcal{A}_*$, hence why we have chosen the form of the Fisher Metric in (\ref{Fisher Metric on A star})}

\begin{equation} \label{Fisher Metric on A star}
    g_{p_A}(U,V) = \text{Tr}_{\text{Vol}_A}\left(\frac{UV}{p_A}\right) = \int_{S_A} \text{Vol}_A(x) \frac{U(x) V(x)}{p_A(x)}
\end{equation}
Here $U,V: S_A \rightarrow \mathbb{R}$ are elements of $T_{p_A}\mathcal{A}_*$, which, in the vein of (\ref{Tangent to Space of Densities}), we identify with measurable functions on the sample space which have zero weight when integrated over the sample space with respect to the reference measure. This guarantees that such random variables can be identified with perturbations to a probability density which maintain the integral normalization condition. 

With all of this terminology in place, we are now prepared to understand the implication of the Petz Map for commutative algebras. To begin, let us remark that the Petz map, as a channel, $\mathcal{P}_{p_A, \mathcal{E}}:\mathcal{B}_* \rightarrow \mathcal{A}_*$, can be associated with a stochastic map $q_{A \mid B}: S_A \times S_B \rightarrow \mathbb{R}$ with

\begin{equation}
    \mathcal{P}_{p_A, \mathcal{E}}(p_B) = \int_{S_B} \text{Vol}_B(y) q_{A \mid B}(x \mid y) p_B(y) \in \mathcal{A}_*
\end{equation}
Thus, the adjoint condition:

\begin{equation}
    g_{p_A}(U, \mathcal{P}_{p_A, \mathcal{E}}(V)) = g_{\mathcal{E}(p_A)}(\mathcal{E}(U), V)
\end{equation}
implies the following. First, the left hand side can be written:

\begin{flalign}
    g_{p_A}(U, \mathcal{P}_{p_A, \mathcal{E}}(V)) &= \int_{S_A} \text{Vol}_A(x) \frac{U(x) \left(\int_{S_B} \text{Vol}_{B}(y) q_{A \mid B}(x \mid y) V(y)\right)}{p_A(x)} \\
    &= \int_{S_A \times S_B} \text{Vol}_A(x) \wedge \text{Vol}_B(y) \frac{U(x) q_{A \mid B}(x \mid y) V(y)}{p_A(x)}
\end{flalign}
Similarly, the right hand side is of the form:

\begin{flalign}
    g_{\mathcal{E}(p_A)}(\mathcal{E}(U),V) &= \int_{S_B} \text{Vol}_B(y) \frac{\left(\int_{S_A} \text{Vol}_A(x) p_{B \mid A}(y \mid x) U(x)\right) V(y)}{\int_{S_A} \text{Vol}_A(x') p_{B \mid A}(y \mid x') p_A(x')} \\
    &= \int_{S_A \times S_B} \text{Vol}_A(x) \wedge \text{Vol}_B(y) \frac{U(x) p_{B \mid A}(y \mid x) V(y)}{\int_{S_A} \text{Vol}_A(x') p_{B \mid A}(y \mid x') p_A(x')}
\end{flalign}
Equating the left hand side and the right hand side for arbitrary functions $U: S_A \rightarrow \mathbb{R}$ and $V: S_B \rightarrow \mathbb{R}$ we therefore find:

\begin{equation}
    \frac{q_{A \mid B}(x \mid y)}{p_A(x)} = \frac{p_{B \mid A}(y \mid x)}{\int_{S_A} \text{Vol}_A(x') p_{B \mid A}(y \mid x') p_A(x')}
\end{equation}
This is Bayes' Law.

\end{appendix}


\end{document}